\documentclass[acmsmall, screen, nonacm]{acmart}

\usepackage{multirow}
\usepackage{subcaption}
\usepackage{enumitem}
\usepackage{tikz}
\usetikzlibrary{automata,positioning}

\newlength\my
\newlength\mx
\tikzset{%
    mynode/.style={circle,draw=black,fill=black!10,inner sep=0pt,minimum size=3.5mm},
    myedge/.style={font=\scriptsize}
}

\theoremstyle{acmplain}

\begin{document}

\title{From Research to Proof-of-Concept: Analysis of a Deployment of FPGAs on a Commercial Search Engine}
\subtitle{White Paper}

\author{Fabio Maschi}
\email{fabio.maschi@inf.ethz.ch}
\orcid{0003-1576-417X}
\author{Gustavo Alonso}
\email{alonso@inf.ethz.ch}
\affiliation{%
  \institution{Systems Group, Department of Computer Science, ETH~Zurich}
  \streetaddress{Stampfenbachstrasse 114}
  \city{Zurich}
  \country{Switzerland}
  \postcode{8092}}

\author{Anthony Hock-Koon}
\email{anthony.hock-koon@amadeus.com}
\author{Nicolas Bondoux}
\email{nbondoux@amadeus.com}
\author{Teddy Roy}
\email{teddy.roy@amadeus.com}
\author{Mourad Boudia}
\email{mourad.boudia@amadeus.com}
\author{Matteo Casalino}
\email{matteo.casalino@amadeus.com}
\affiliation{%
  \institution{Amadeus}
  \city{Sophia-Antipolis}
  \country{France}}

\renewcommand{\shortauthors}{Maschi et al.}




\begin{abstract}
FPGAs are quickly becoming available in data centres and in the cloud as a one more heterogeneous processing element complementing CPUs and GPUs. There are many reports in the research literature showing the potential for FPGAs to accelerate a wide variety of algorithms, which combined with their growing availability, would seem to also indicate a widespread use in many applications. Unfortunately, there is not much published research exploring what it takes to integrate an FPGA into an existing application in a cost-effective way and keeping the algorithmic performance advantages. Building on recent results exploring how to employ FPGAs to improve the search engines used in the travel industry, this paper analyses the end-to-end performance of the search engine when using FPGAs, as well as the necessary changes to the software and the cost of such deployments. The results provide important insights on current FPGA deployments and what needs to be done to make FPGAs more widely used. For instance, the large potential performance gains provided by an FPGA are greatly diminished in practice if the application cannot submit request in the most optimal way for the FPGA, something that is not always possible and might require significant changes to the application. Similarly, some existing cloud deployments turn out to use a very imbalanced architecture: a powerful FPGA connected to a not so powerful CPU. The result is that the CPU cannot generate enough load for the FPGA, which potentially eliminates all performance gains and might even result in a more expensive system. In this paper, we report on an extensive study and development effort to incorporate FPGAs into a search engine and analyse the issues encountered and their practical impact. We expect that these results will inform the development and deployment of FPGAs in the future by providing important insights on the end-to-end integration of FPGAs within existing systems. \end{abstract}

\maketitle


\section{Introduction}

FPGAs are becoming pervasive in data centres and cloud platforms and have been made available to cloud users as one more option within an increasingly heterogeneous computing landscape. Yet, in spite of the abundant literature demonstrating the performance advantages, architectural improvements, energy savings, etc., that can be achieved with an FPGA, taking advantage of all possibilities FPGAs have to offer for existing applications is less than straightforward. There are several reasons for this. Among them, the fact that FPGAs are a very novel component  in the context of conventional computing and there is not enough experience on what designs work best. A prime example of a very successful deployment of FPGAs in the cloud is Microsoft's Catapult. The engineers behind the design have been very open and have described the process in several papers~\cite{putnam2014catapult, caulfield2016, Firestone2018} that illustrate the substantial changes in the system from version to version and the different use cases considered until one final architectural configuration and a concrete use case was found that justified the investment and the deployment at such scale.

Unlike what has been done for many other hardware and software systems over the years, there are still very few studies analysing FPGA deployments and what needs to be done to make them cost-efficient in practice. While some works present the hardware efficiency \emph{vis-à-vis} the cloud infrastructure~\cite{DBLP:conf/cloudcom/FahmyVS15, DBLP:conf/cf/ChenSZWFCW14, DBLP:conf/fccm/BymaSBLC14}, and the vast FPGA literature emphasises kernel acceleration under a stand-alone context, very few analyse the deployment efficiency of applying FPGAs in real computing systems. In this paper, we present one such study, based on recent research results that demonstrated significant potential gains when implementing part of a search engine on an FPGA~\cite{maschi2020,Mohsen20}. The initial prototype was turned into a Proof-of-Concept deployment tested on the real system and computing infrastructure, as well as evaluated against real data and under the constraints imposed by the existing search engine. The exercise delivered many important insights regarding: (i) the integration of FPGAs into existing software stacks; (ii) the adequate balance between CPU and FPGA; (iii) issues of modularity, interfaces, and request management; (iv) the overheads of making the system feature complete; and (v) overall cost, which turned out to be quite different than originally expected. In what follows, we report on the concrete steps undertaken to turn the research idea into a real component, and provide an extensive evaluation of the resulting performance and estimated costs of the possible deployments. Along the way, details are provided regarding the design decisions and the constraints imposed by the existing search engine.

The starting point for the study is a use case from the airline industry: the flight search engine used by Amadeus to provide search services to travel companies. Like many other search engines, Amadeus' flight search engine is a large scale distributed system comprising many different components. The \emph{Minimum Connection Time} (MCT) module is one such component and it is implemented atop a Business Rule Management System (BRMS). It is used in the early stages of the search 
and plays a key role in terms of the performance and total cost of operating the search engine. When a query looking for flights needs to be processed, a large number of potential routes have to be computed. For all routes that are non-direct flights, the MCT module is invoked to ascertain the minimum connecting time to the next flight. Thus, it needs to fulfil stringent performance requirements on both latency per query and overall query throughput. Because it is used as a module within a larger system, a common architecture for Decision Rule Engines~\cite{DBLP:journals/cea/Mazon-OlivoHMP18}, there are additional constraints in terms of the amount of memory used that determine what type of BRMS can be employed in practice. In the current deployment, the MCT module is responsible for 40\% of the computing resources used by the Domain Explorer, making it an ideal starting point to optimise the system.

An initial implementation of MCT on top of an FPGA proved to be a significant improvement over the existing system along several dimensions~\cite{maschi2020}. By using a Non-deterministic Finite State Automaton (NFA) and exploiting the inherent parallelism and pipeline possibilities available on the FPGA, the performance gains over the existing system were impressive. \textsc{erbium}\footnote{\textsc{ERBium} is open source: \url{github.com/fpgasystems/erbium}}, the hardware engine, processes queries three orders of magnitude faster than the fastest CPU implementation. A single instance delivers a throughput of up to 50 million queries per second, 800 times larger than a single CPU instance. In theory, with such throughput, a single instance would be able to deal with the entire MCT workload of the search engine. In terms of cost, it reaches 60 billion queries per U.S. Dollar when running in the cloud (Amazon AWS), 20 times cheaper than the most cost-efficient CPU deployment. The downtime for updating the rules is only 500~$\mu$s, four orders of magnitude faster than in the current system, which can be used to improve the overall availability of the search engine. Moreover, the performance gains indicated that the improvements could be exploited in several ways in addition to increasing throughput and reducing latency: the quality of the search could be improved by considering more options as part of the search; the computing capacity needed for the search engine could be reduced; and the architecture became more flexible, offering several ways to integrate the MCT and other components (see~\cite{maschi2020,Mohsen20} for more details and an extensive discussion of the deployment possibilities of FPGAs within the flight search engine).

Based on these results, a decision was made to integrate the FPGA-based MCT module into the actual search engine and evaluate its performance in a real setting. In the following sections, we report on this exercise and discuss the insights gained in the process. First, the initial implementation was based on a now obsolete model of the rules used in MCT. These rules are standardised and a new version of the standard has become recently available requiring quite a few changes to the FPGA design. These changes reflect in many cases the differences between the kind of simplified design often found in the research literature and a feature-complete component, so we discuss the impact that such changes had on the overall system design (Section~\ref{sec:hardwaredesign}). Second, the MCT module is just one component of a larger engine with the rest of the system remaining in the CPU. When a module is migrated from being a software module to become a component running on the FPGA, there are several options to integrate it in the overall design. These options have a significant impact on performance and need to be carefully evaluated (Section~\ref{sec:systemintegration}). Third, the complete system needs to be tested in an environment where the rest of the components are also active and might potentially result in an infra-utilisation of the FPGA, something we observe in practice and document with experiments (Section \ref{sec:business_logic}). Finally, one of the purposes of the original exercise was to explore ways to improve the performance of the search engine when running in the cloud (Amazon AWS). The initial results indicated that the FPGA design could potentially reduce the number of computing nodes needed, thereby resulting in a lower deployment cost. When we analysed the Proof-of-Concept system resulting from all the changes just described and considering the restrictions imposed by both the cloud platforms and the current search engine, a more differentiated picture appeared. Depending on several factors, using an FPGA would go from reducing the cost in a substantial manner to resulting in a much more expensive system with limited overall gains. The reasons for such apparently bewildering behaviour are complex and we explain them in detail in Section \ref{sec:systemdeployment}.

Taken together, these insights provide a very accurate picture of the challenges met when trying to deploy an FPGA in an existing system. By describing in detail the interactions between the software and the FPGA, the use case we explore can help inform further developments regarding FPGA deployments in the cloud as several of the issues we have encountered are structural limitations that will arise in other systems as well. Our results indicate that successful FPGA deployments are likely to require a careful software-hardware co-design of the entire system, with approaches that only consider the FPGA side of the system being limited by many of the same design considerations we discuss in this paper. 


\section{Background}

In this section, we cover necessary background information on the Amadeus flight search engine, the Travel Solution concept, and the Minimum Connection Time module. We also introduce the notation that will be used in the rest of the paper.


\subsection{A Flight Search Engine}

\begin{figure}[t]
  \centering
  \includegraphics[width=0.7\linewidth]{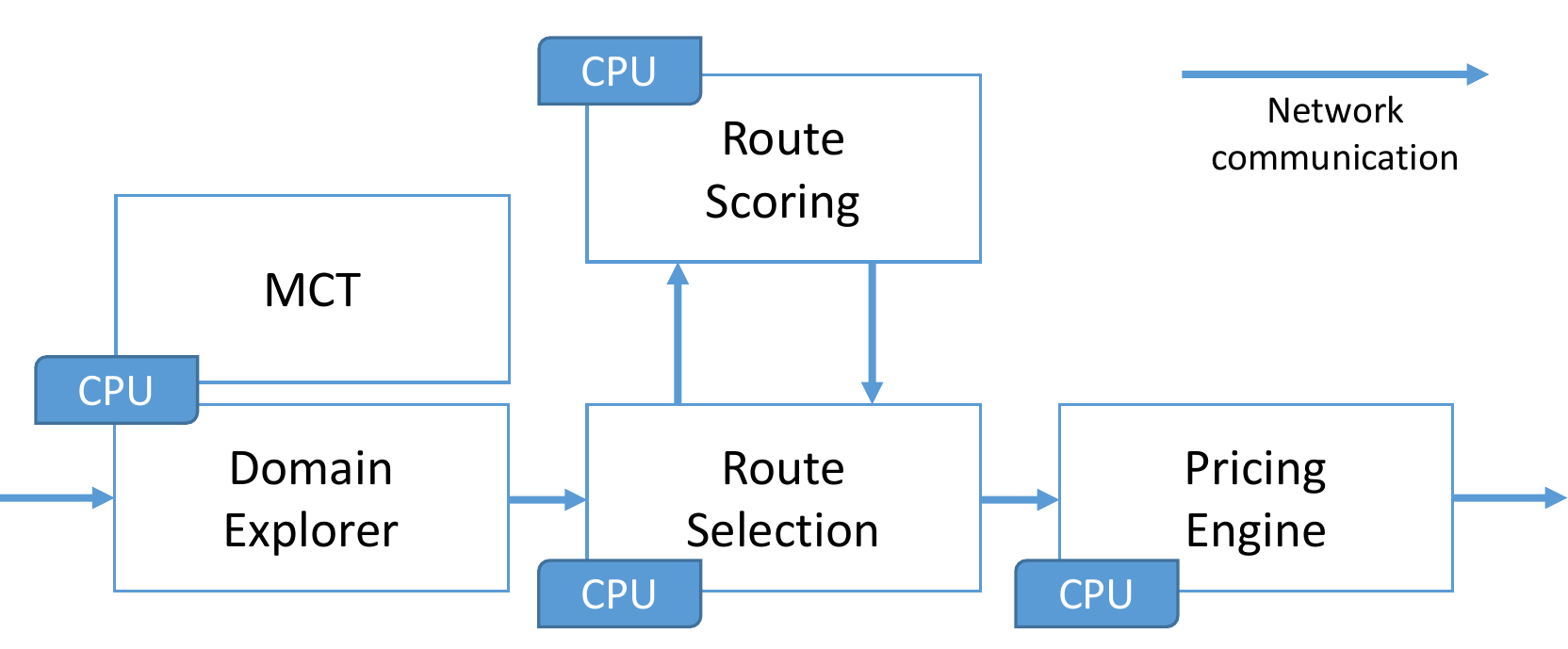}
  \caption{Modular depiction of the Flight Search Engine used by Amadeus.}
  \label{fig:flightsearchengine}
\end{figure}

The Flight Availability Search and Pricing Engine (Figure~\ref{fig:flightsearchengine}) is a search engine that, given a query (here called \emph{user query}) specifying an origin, a destination airport, and the corresponding dates, responds with a number of potential routes between the departure and arrival points with the corresponding flight information and prices. The engine works as an online interactive service and has strict Service Level Agreements (SLAs) in terms of response time and throughput, as it is used by many companies to provide travel services to end customers. Each component has a corresponding latency bound that must be met, so that the overall query response time remains under four seconds. Answering queries may require to explore a very large number of options, e.g., when a user is flexible on the departure and/or returning dates and the route involves several legs.

The engine is divided into several components. The \emph{Domain Explorer} searches the flight domain, exploring all possible connecting airports, carriers and flights combinations; and pre-selects a number of potential routes. These routes are fed into the \emph{Route Selection} component, which uses heuristics and a decision tree ensemble~\cite{Mohsen20} (module \emph{Route Scoring}) to reduce further the set of potential routes most likely to be bought. The reduced set is forwarded to the \emph{Pricing Engine}, which then computes the price for each flight combination. The engine is currently implemented in a data centre architecture, where the different components run on different machines, and many machines are used to run the components in parallel to reach the necessary throughput. The system is large, with each one of the components using several hundred machines. In recent work, we have explored the complex interplay between the number of routes being considered and the overall throughput/latency constraints, and we refer to~\cite{Mohsen20} for more details on that aspect of the system and how we successfully used hardware acceleration to optimise the Route Scoring module.

The MCT module is part of the Domain Explorer and runs on the same machines. This design is necessary to cut on the number of network hops needed to process a query, but imposes restrictions on the rule engine. For instance, Drools~\cite{Drools}, the business rule engine commonly used by Amadeus, requires 9.5~GiB of main memory to run the hundreds of thousands of MCT rules involved. Such a high resource consumption makes it impossible to embed it within the Domain Explorer given the existing architecture. Deploying it stand-alone in its own set of servers is not an option either, as it would cause network overhead. As a result, the MCT is a customised implementation in C++ intended to provide the required performance while minimising memory consumption. In addition, being part of an online interactive service as the Domain Explorer, the downtime of the MCT module for rule set updates is restricted to a minimum. Nonetheless, the current MCT module deployment consumes 40\% of the computing resources allocated to the Domain Explorer, making it an obvious starting point to make the system more efficient.

\subsection{Travel Solutions}
In the early stages of the flight search engine, for a given user query, several routes connecting the origin and destination airports are explored. Each combination of such single-direction routes (e.g., from Nice to Paris, from Paris to Boston), carriers (e.g., Air France and Delta), flight numbers (e.g., AF7701 and DL8604), and flight times is called a \textit{Travel Solution} (TS). If a direct flight exists, it will result in a single TS; alternative indirect routes are captured in an additional TS each. Given the search space of the routes and the response time constraints for the search engine, the current system has been tailored to explore up to 1,500 TS's per user query, and the number of connecting airports within the same TS is limited to five. Special user queries may trigger the evaluation of more TS's, but these are a minority workload. The number of TS's considered per user query plays an important role when integrating the FPGA, as it will be discussed later. 


\subsection{MCT: Filtering Impossible Connections}

The rules determining the minimum connection time are provided by each airline and are standardised by IATA~\cite{ssim2020, mctv1, mctv2}. The connection time is affected by a number of variables (e.g., airports, terminals of arrival and departure, whether passport or immigration controls are involved, aeroplane model, time of day, etc.). The rules defining the MCT change regularly, so airlines can adapt their flight offers to their most recent logistic and commercial constraints (e.g., temporal peaks in flow of flights or passengers, changes in connection preferences between airlines, construction work at airports, etc.). The flight search engine encompasses all airports worldwide and every airline contributes a long list of rules for every airport where they operate. The current version of the MCT module operates on over 160k rules. Table~\ref{tab:rules} shows a simplified, but syntactically representative example of how the MCT rules look like (the actual rules have thirty-four criteria). We refer to~\cite{maschi2020} for further information on how the MCT module processes the rules.

\begin{table}
  \centering
  \caption{Example of the rules $r_{[0,5]}$ determining the Minimum Connection Time (actual rules have 34 criteria), and a possible query $\rho_0$.}
  \setlength{\tabcolsep}{3pt} 
  \renewcommand{\arraystretch}{1.35}
  {\footnotesize
  \begin{tabular}{@{}lllll|ll@{}} \toprule
  & Airport & Time frame & Region & Terminal & Decision & Precision \\ \midrule
  $r_0$ & ZRH & * & International & * & 90 min & Low \\
  $r_1$ & ZRH & * & Schengen & T1 & 25 min & Middle \\
  $r_2$ & ZRH & Summer '21 & Schengen & T1 & 40 min & High \\
  $r_3$ & ZRH & Winter & Schengen & T1 & 25 min & High \\
  $r_4$ & CDG & Winter & Schengen & T1 & 25 min & High \\
  $r_5$ & CDG & Sundays & International & T2 & 45 min & High \\ \midrule
  $\rho_0$ & ZRH & 12\textsuperscript{th} Aug '21 & Schengen & T1 &  \\ \bottomrule
  \end{tabular}}
  \label{tab:rules}
\end{table}

\section{Hardware Design}
\label{sec:hardwaredesign}

In what follows, we describe the operational characteristics of the FPGA-based engine, \textsc{erbium}, and the additional modifications made to its previous version to fully support the MCT and accommodate a new standard used to specify the rules.

\subsection{\textsc{erbium} engine}

\begin{figure}
  \centering
  \includegraphics[width=0.9\linewidth]{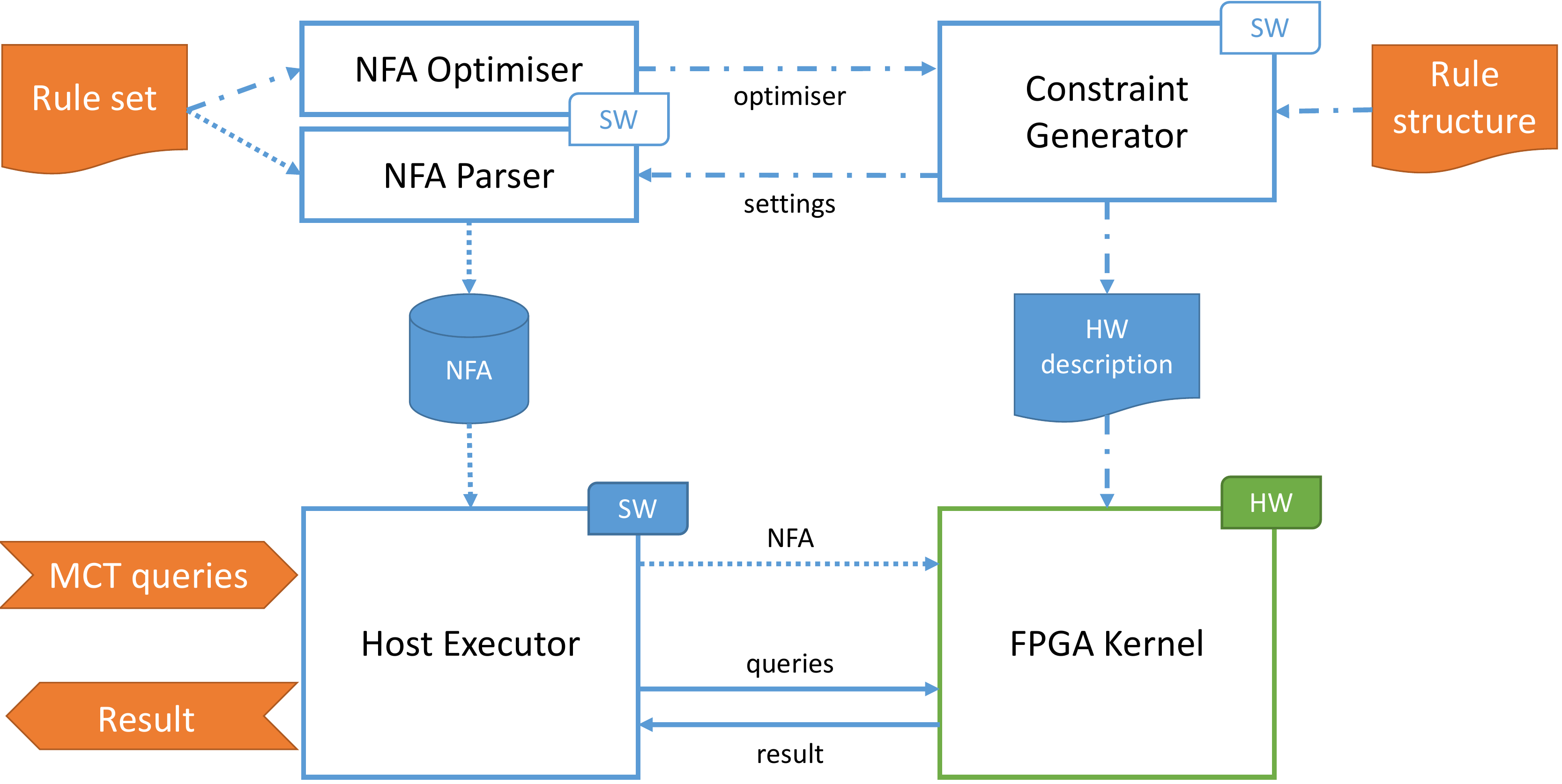}
  \caption{The \textsc{erbium} engine overview. In orange, the external inputs and outputs; in blue, the CPU modules; and in green the FPGA kernel. Dashed lines represent processes that are conducted offline.}
  \label{fig:erbium}
\end{figure}

Figure~\ref{fig:erbium} depicts \textsc{erbium}~\cite{maschi2020}, the NFA-based Business Rule Engine hardware accelerator. Its different elements can be decomposed into offline and online modules. In the first group, the \emph{NFA Optimiser} uses statistical heuristics on the rule set to optimise the NFA shape (i.e., the order of the criteria within the NFA) for both memory and latency requirements. The \emph{Constraint Generator} customises the hardware kernel according to the rule structure (i.e., number of criteria, data-types, operators) and the NFA shape. The \emph{NFA Parser} builds the NFA memory file based on the current hardware settings and on the rule set. These three modules run in centralised machines of the cluster and are used to generate updated versions of the NFA every time the rules change, or a more optimised version of the hardware engine when needed.

In the second group of components, the \emph{Host Executor} is responsible for loading the NFA data into the FPGA internal memory, as well as for sending the MCT queries and fetching their results, which is the main processing flow. The \emph{FPGA Kernel} is the hardware accelerator. These two modules compose the computing nodes of the system, and can be replicated as many times as needed to increase the throughput of MCT queries. We refer to~\cite{maschi2020} for more details on the interface between Host and Kernel, as well as their communication protocols constraints.

External inputs and outputs to the \textsc{erbium} engine have different time persistence. The Rule structure can be seen as the schema of a table in a classic database, and therefore can be considered as static information per use case. The Rule set is updated once a day with information provided by the airlines. The daily updates do not significantly change the statistics of the data, hence the optimisation results from one day are likely valid for a considerable period. Combined with the static rule structure, the hardware description is rarely changed. As a consequence, the downtime required by the reconfiguration of the FPGA to install a new version of the rules is negligible.

\subsection{Adapting from MCT v1 to v2}

The rules used in MCT are expressed using a standard formalised by the International Air Transport Association (IATA). The first integration challenge arose from a new version of the standard initiated by IATA~\cite{mctv2}. In a nutshell, the rules considered in  MCT v1 were a simple conjunction of independent predicates over the attributes of the rules. The rule structure of the new version of the standard introduces inter-dependencies among the criteria, as well as a different priority weight hierarchy. It is worth mentioning that these new features are not yet used by the airlines to describe their current rules, but we wanted the system to be ready for it when the standard is fully adopted.

Such modifications represent, from a software point of view, additional code to check different conditions and additional variables in the rule structure. In contrast, for a hardware engine on an FPGA, the behaviour of the system needs to be adapted, which has further consequences. In what follows, we describe the four main changes required to support MCT v2, as illustrated in Figure~\ref{fig:nfaoptimisations}. These changes are illustrative of the challenges faced when deploying a prototype in a real system. Also, the prototype needs to be adapted to the actual data representation used instead of using formats easier to handle in an FPGA.



\subsubsection{Criteria merging} 
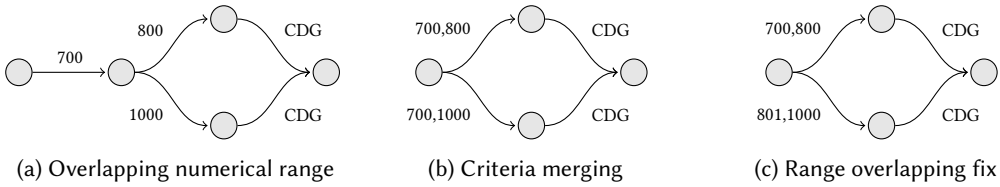
\begin{figure*}[t]
    \centering
    \begin{subfigure}[t]{0.33\textwidth}
        \centering
        \begin{tikzpicture}[shorten >=1pt,node distance=\mx, auto]
    \node[mynode] (q_0) {};
    \node[mynode] (q_700) [right=of q_0] {};
    \node[mynode] (q_1k)  [right=of q_700, yshift=-0.5\my] {};
    \node[mynode] (q_800) [right=of q_700, yshift=+0.5\my] {};
    \node[mynode] (q_cdg) [right=of q_800, yshift=-0.5\my] {};
    
    \path[->, out=0, in=180]
    (q_0)   edge node[myedge]       {700} (q_700)
    (q_700) edge node[myedge, swap] {1000} (q_1k)
            edge node[myedge]       {800} (q_800)
    (q_800) edge node[myedge]       {CDG} (q_cdg)
    (q_1k)  edge node[myedge,swap]  {CDG} (q_cdg);
\end{tikzpicture}
        \caption{Overlapping numerical range}
    \end{subfigure}%
    ~ 
    \begin{subfigure}[t]{0.33\textwidth}
        \centering
        \begin{tikzpicture}[shorten >=1pt,node distance=\mx, auto]
    \node[mynode] (q_0) {};
    \node[mynode] (q_710) [right=of q_0, yshift=-0.5\my] {};
    \node[mynode] (q_780) [right=of q_0, yshift=+0.5\my] {};
    \node[mynode] (q_cdg) [right=of q_780, yshift=-0.5\my] {};
    \path[->, out=0, in=180] 
    (q_0)   edge node[myedge, swap] {700,1000} (q_710)
            edge node[myedge]       {700,800}  (q_780)
    (q_780) edge node[myedge]       {CDG} (q_cdg)
    (q_710) edge node[myedge,swap]  {CDG} (q_cdg);
\end{tikzpicture}
        \caption{Criteria merging}
        \label{fig:nfaoptimisations_merging}
    \end{subfigure}
    ~
    \begin{subfigure}[t]{0.33\textwidth}
        \centering
        \begin{tikzpicture}[shorten >=1pt,node distance=\mx, auto]
    \node[mynode] (q_0) {};
    \node[mynode] (q_710) [right=of q_0, yshift=-0.5\my] {};
    \node[mynode] (q_780) [right=of q_0, yshift=+0.5\my] {};
    \node[mynode] (q_cdg) [right=of q_780, yshift=-0.5\my] {};
    \path[->, out=0, in=180] 
    (q_0)   edge node[myedge,swap] {801,1000} (q_710)
            edge node[myedge]      {700,800} (q_780)
    (q_780) edge node[myedge]      {CDG} (q_cdg)
    (q_710) edge node[myedge,swap] {CDG} (q_cdg);
\end{tikzpicture}
        \caption{Range overlapping fix}
    \end{subfigure}
    \caption{Transformations on the NFA model to support additional features. (a) a simplified version of the NFA covering the range values [700,1000] and [700,800] to CDG airport; (b) range \emph{start} and \emph{end} values are merged as a pair of values; (c) overlapping ranges are eliminated, so the most precise range is unique as a match.}
    \label{fig:nfaoptimisations}
\end{figure*}

MCT v2 expands all numerical range criteria from a pair-of-values to two independent criteria. These changes can be considered as a pure syntactic modification (turning a complex attribute storing a numeric range into two separate criteria with the minimum and the maximum of the range~\cite{HirzelSSGG13}). As such, they just require to change the \emph{NFA Parser} module, which now encodes range based criterion accordingly. This type of transformation is easy to capture in the NFA structure and is not visible to the outside as it only changes the number of steps in the NFA. The original design used a step for each attribute considered. The new format increases the number of attributes/criteria and, thus, results in a bigger NFA with additional steps. In addition, changing the number of steps results in the cardinality of the corresponding attribute varying significantly. For instance, a criteria merging operation removes one NFA stage, but the cardinality of the new attribute is now the Cartesian product of the values of the original attributes, as illustrated in Figure~\ref{fig:nfaoptimisations_merging}. The cardinality at each stage has a direct impact on both the memory required to store the NFA transitions and the latency needed to traverse the graph, a metric directly connected to the response time of a query. Thus, while the new rules do not require to change the design on the FPGA, the resulting larger NFA places additional constraints on resources and affects latency results. 


\subsubsection{Precision weight for ranges}
Each rule has a priority weight that models how precise or generic the rule is. This precision weight is used at runtime for choosing the most precise rule among all the ones that have matched the query. The use of priority weights makes it easier not only to answer queries when there is not much information, but also to make the answer more precise if the query contains more information. In MCT v1, every criterion has its intrinsic and unique weight value, and the precision weight of a rule is the sum of the precision weights associated to the different criteria used by the rule. In other words, a rule whose criterion value is a wildcard has a precision weight reduced by the intrinsic weight related to that criterion.

MCT v2 introduces an additional layer of priority management depending on the flight number range size. Larger ranges are less precise, and therefore carry less precision weight than a shorter one. For range criteria, the precision weight is then a combination of the intrinsic weight and a dynamic one coming from the range value used in the rule. Based on this, there are two options to address this new requirement: (i) modifying the hardware engine to support the additional precision layer, and (ii) modelling this layer within the NFA generation~\cite{DBLP:conf/birthday/IlieNY04}.

The former option would require a major redesign to introduce the necessary computation of the priority. The latter option does not require a redesign but, as above, results in a larger and more involved NFA. We opt for using the \emph{NFA Parser}. This module is able to detect rules with overlapping ranges offline. It can then translate the ranges into new, non-overlapping rules, ensuring that a particular flight number can match only one rule. On the one hand, by doing so, we delegate the computing cost that would be otherwise done online to the offline module. On the other hand, the number of rules increases, resulting in a bigger memory requirement.

As we will see later, increasing the number of rules ---~and, thus, of the NFA~--- is not (yet) an issue for two reasons: the inherent performance of the rule engine on the FPGA is currently limited by the PCIe bandwidth and the ability to feed big enough batches (see below); and the daily rule sets do not present substantial overlapping flight number ranges. 
Currently, the number of additional rules due to conflicting ranges varies between zero to a few hundred among an average of 160k rules, so the resulting increase in size is quite moderate. 

\subsubsection{Cross-matching criteria}
The initial version of the MCT rule structure was composed of independent criteria. Every query criterion value maps to a single rule criterion (e.g., the query value ``CDG'' would map to ``Airport'' only). The new version, however, introduces the concept of cross-matching criteria. In certain cases, a single criterion input value can be matched against two criterion rule values, depending on the value of a third criterion. Consider the following example. In the travel industry, it is necessary to differentiate between a flight commercialised by a carrier (marketing carrier), but operated by a second carrier (operating carrier). This is called a \emph{code-share} flight. IATA expanded a single criterion \emph{Arrival Carrier} to three to support this use case: a first that represents the carrier that commercialised the flight, the marketing carrier; a second that represents the operating carrier, and a third that indicates the code-share situation. In the case the code-share indicator is false, the operating carrier value from the query should match the marketing carrier value of the rule, requiring a cross-matching criterion operation.

In order to support this new functional behaviour, important changes need to be done on the hardware kernel side, so we decided to re-model the logic behind it and encapsulate it on the NFA. Two criteria are used: (i) marketing, and (ii) operating carrier. In the case the code-share indicator is present, the rule remains as it is declared. In the case there is no code-share indicator, the marketing and operating carrier are the same, therefore we duplicate the value to both criteria. This feature can be modelled within the NFA generation. Unlike the previous two cases, the change typically does not induce any major changes to the latency, memory requirements nor, more importantly, on the precision weight structure. 


\subsubsection{Flight number and code-share flight number}
Similarly to marketing and operating carrier, the flight number criteria were adapted to support code-share flights. In MCT v2, a single field \emph{flight number range} is provided by the rule, and its value should be either matched against the operating carrier flight number, or against the marketing carrier flight number, according to the code-share indicator. 
We support this new constraint by adding a criterion to the rule, uniquely representing the code-share flight number range. Its value is based on the rule context and is populated whenever the code-share flight range is the one that must be matched. 
Just as for the carrier values, the MCT query provides both (marketing and operating) flight numbers, and they are matched against the correct value according to the logic modelled in the NFA by the \emph{NFA Parser}.

These optimisations allow the hardware engine to remain intact, as it does not violate the initial constraint requiring that rules are a conjunction of independent criteria. The expressiveness of the rules that do not violate this constraint is limited, and must be taken into account when envisaging long-term deployments. For the MCT use case, however, the expressiveness supported by the NFA model used in \textsc{erbium} is enough to express all the rules.

\subsection{Experimental Evaluation}

\begin{figure*}
    \centering
    \begin{subfigure}[t]{0.50\textwidth}
        \centering
        \includegraphics[width=\linewidth]{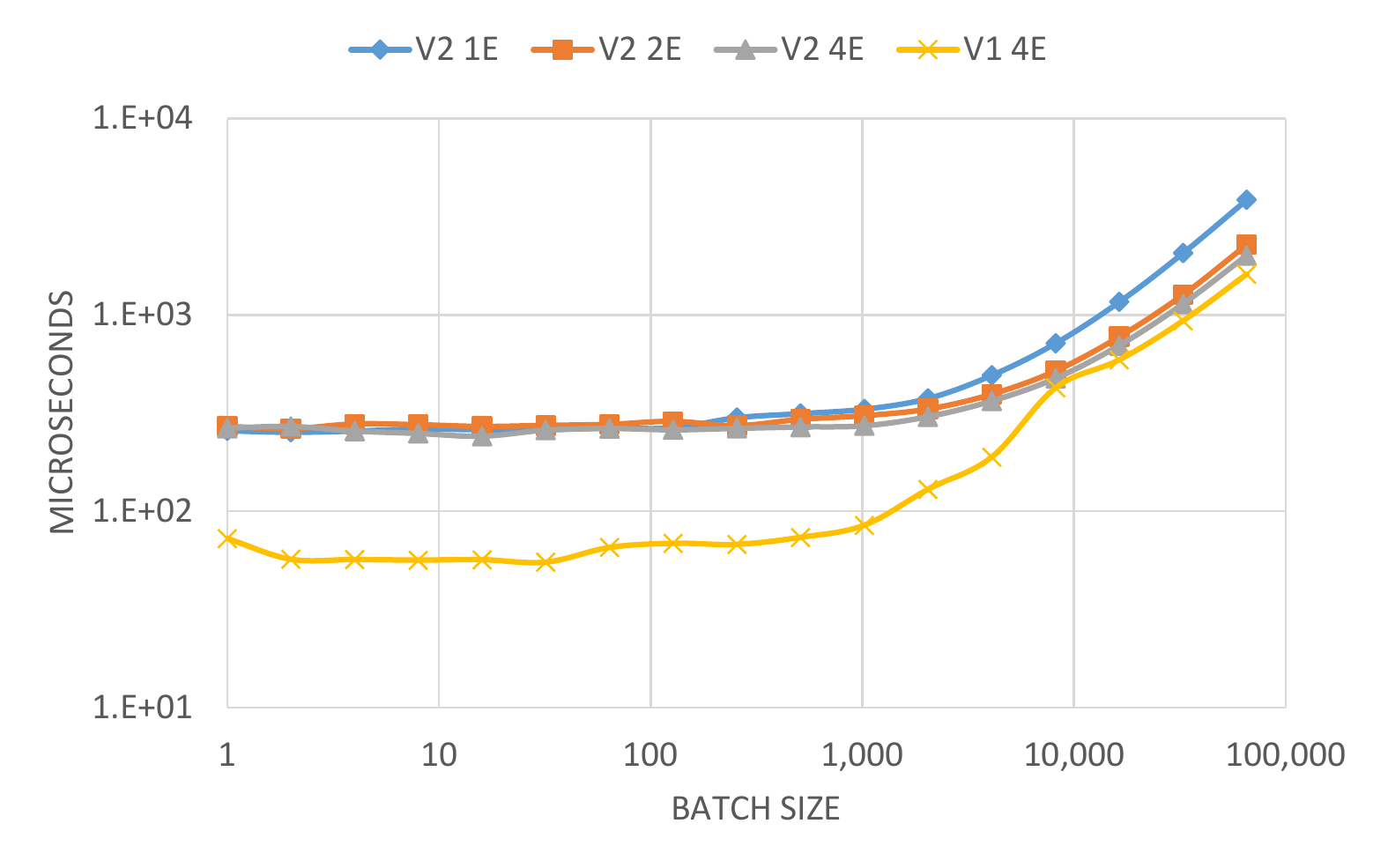}
        \caption{Execution Time}
    \end{subfigure}
    ~ 
    \begin{subfigure}[t]{0.50\textwidth}
        \centering
        \includegraphics[width=\linewidth]{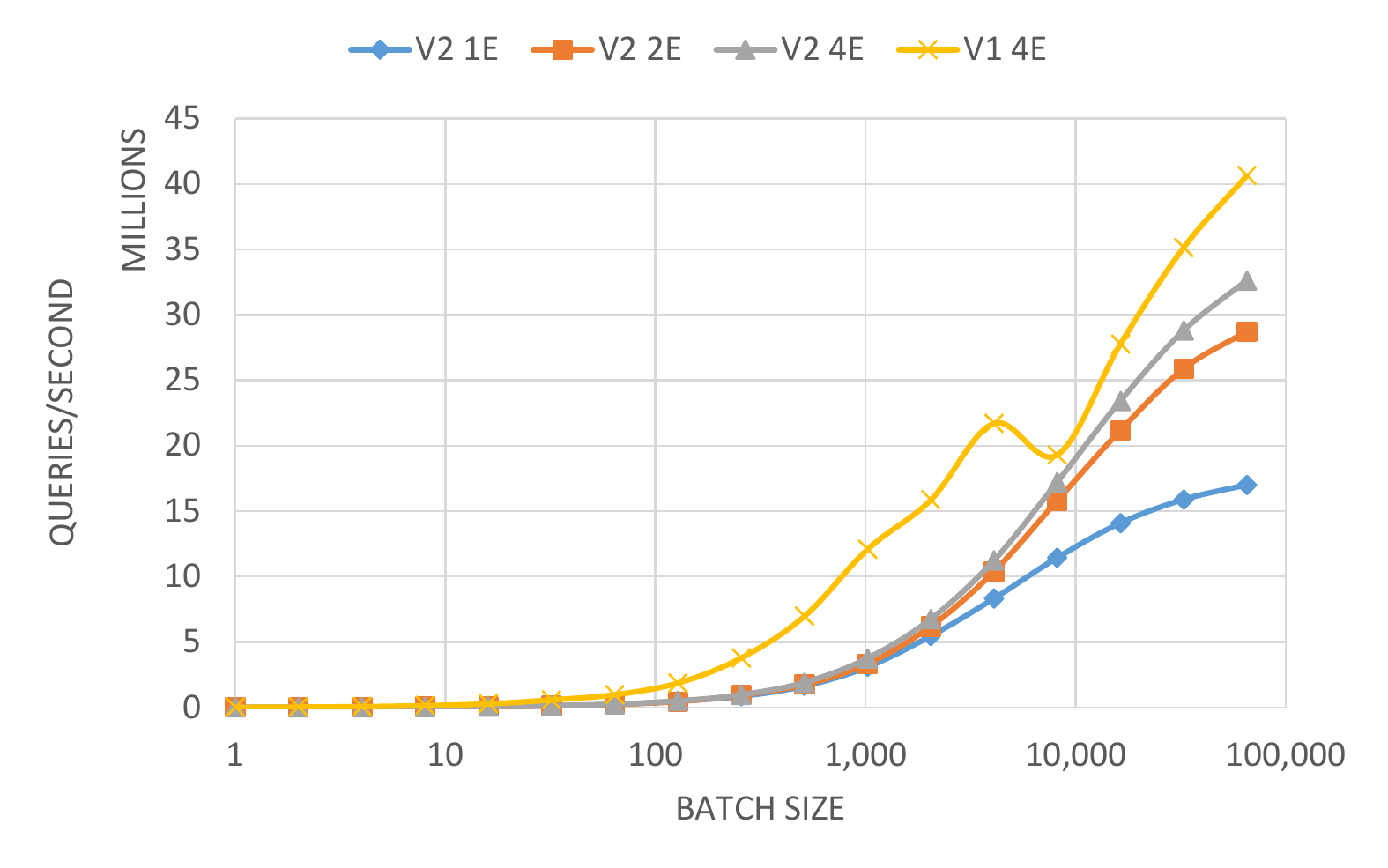}
        \caption{Throughput}
    \end{subfigure}
    \caption{Execution time in microseconds, and throughput in MCT queries per second as a function of the batch size. Stand-alone hardware engine processing MCT v1 (yellow) and MCT v2 queries (blue, orange and grey). The v2 experiments were conducted in AWS F1 instances.}
    \label{fig:mctv2benchmark}
\end{figure*}

We benchmark a stand-alone deployment of \textsc{erbium} processing MCT v2 rules after the modifications reported above. The experiments were executed in the cloud (Amazon AWS F1 instance), and \textsc{erbium} was configured using one kernel containing one, two and four NFA Evaluation Engines. We compare those results with the MCT v1 setup using 4 NFA Evaluation Engines on-premises (Alveo U250, QDMA shell) reported previously~\cite{maschi2020}. Per batch size, we compute one thousand travel solutions. All the numbers presented correspond to the 90\textsuperscript{th} percentile of the experiments per batch, as that matches the SLA of the search engine.

Although the rules in MCT v1 and v2 differ in both semantics and rule set, the input data coming from the \emph{Domain Explorer} remains relatively similar. Hence, the performance comparison gives us an interesting perspective of how the changes caused by the new standard used in v2 affect performance. The MCT v2 deployment is 56\% more resource-intensive due to the larger set of rules and more complex rule structure, but requires 4\% less FPGA memory. The reduction in memory consumption is explained by the more homogeneous distribution of NFA transitions throughout the different NFA levels in v2, since it contains a larger number of relevant criteria than v1. In contrast, the NFA depth, and hence the hardware pipeline, is larger; which affects the overall latency. This is a consequence of having 26 consolidated criteria in v2, against only 22 in v1. The main latency drawback for the v2 deployments in the cloud for small batch sizes comes from the FPGA shell difference. The v2 deployment on AWS can only use the XDMA shell, whereas the v1 experiments takes advantage of the streaming interface provided by the QDMA shell~\cite{xilinx-qdma}. The different interfaces play a big role in the latency of small data transfers up to 1,024 queries per batch. Figure~\ref{fig:mctv2benchmark} shows that both engine setups respond similarly as long as their pipeline is not fully saturated, which does not happen with loads smaller than 100k queries per batch. Above this threshold, MCT v1 is able to process up to 40 million queries per second, while MCT v2 saturates at 32 million queries per second, by virtue of a 11\% lower operating frequency caused by the bigger NFA size.

In summary, when looking only at the performance, we observe a significant performance loss over the original design when all the new requirements are considered. This can be seen as illustrative of the difference between a prototype and a system that needs to consider all corner cases. Some of the loss is caused by the larger NFA which affects latency and results in a lower operating frequency. Another part of the loss arises due to differences in the shells available, with the new deployment having to use a blocking interface rather than a streaming one, which also causes additional costs when passing data and extracting the results~\cite{maschi2020}. The performance loss is lower for large batch sizes as the overhead is distributed across more queries. This offers an opportunity to compensate the losses in the new design. 


\subsection{Discussion}

The example illustrates the practical relevance of not only the design on the FPGA, but of the tools related to it, in this case the NFA generator. The search engine will face over time many changes to the rules format, the standards, the criteria considered, etc. This needs to be factored in when moving part of the system to an FPGA. Without a suitable generator, each one of these changes caused by the new standard would have required a complete redesign of the circuit on the FPGA, which would have made the whole approach highly impractical due to the high cost of development and its effects on the life cycle of the search engine. In our experience, the often heard criticism regarding the cost of development on FPGAs refers in reality to the cost of evolving and maintaining the design over the entire lifetime of the system. With the appropriate tools and by adopting flexible designs that can be easily adapted, a significant part of the development cost over time can be reduced to a minimum. In our case, the major changes in the rules standard could be addressed through changes to the \emph{NFA Parser}, a software component that is much easier to maintain and modify, while keeping the core FPGA design virtually identical from version to version. Of course, the drawback is that the NFA implemented on the FPGA needs to remain a generic one, so that it can accommodate changes without affecting its structure.

The trade-off of generality and easy of maintenance at the cost of designs that do not achieve maximum performance is a common one in software, but one that still need to be considered in many FPGA-based systems where performance often trumps all other considerations~\cite{nope}. As our use case illustrates, the practical viability of a design hinges not only on its performance, but also on its ability to easily evolve and adapt to changes. This could be determined by tools outside the FPGA, as demonstrated in our project, where the NFA generator has become a key component that has allowed us to absorb much of the additional complexity without having to completely redesign the circuit on the FPGA.

The performance comparison shows the importance of considering the life cycle of a real system of this scale. In practice, the load on all components tends to grow over time, as the new version of the MCT rules demonstrates. Thus, a design whose performance is maximised for a particular problem size might not be the best option on the medium and long term. Instead, a flexible design is preferred as it facilitates maintenance and evolution. The larger number of rules in version 2 result in a larger NFA which, in turn, affects the frequency that can be reached. However, the impact of the frequency can be compensated to a large extent by using larger batch sizes. The difference in interfaces (streaming vs blocking) among the shells is something that we expect will eventually disappear, bringing the curves closer for all batch sizes. Thus, we consider that the FPGA design has proven to be scalable and can be evolved without a high development cost, since the complexity is absorbed by a software component that is easier to develop and maintain, but that also becomes an intrinsic element in the system. As a lesson learned, when considering a design, it is important to consider how it could be scaled and modified over time. Ultra-optimised designs that need to be redesigned from scratch to accommodate changes are unlikely to pay off in real deployments. 




\section{System Integration}
\label{sec:systemintegration}

In this section, we discuss the integration of \textsc{erbium} into the complete flight search engine and how assumptions made in software and existing interfaces impact the resulting performance.

\subsection{Setup}

\begin{figure}
  \centering
  \includegraphics[width=\linewidth]{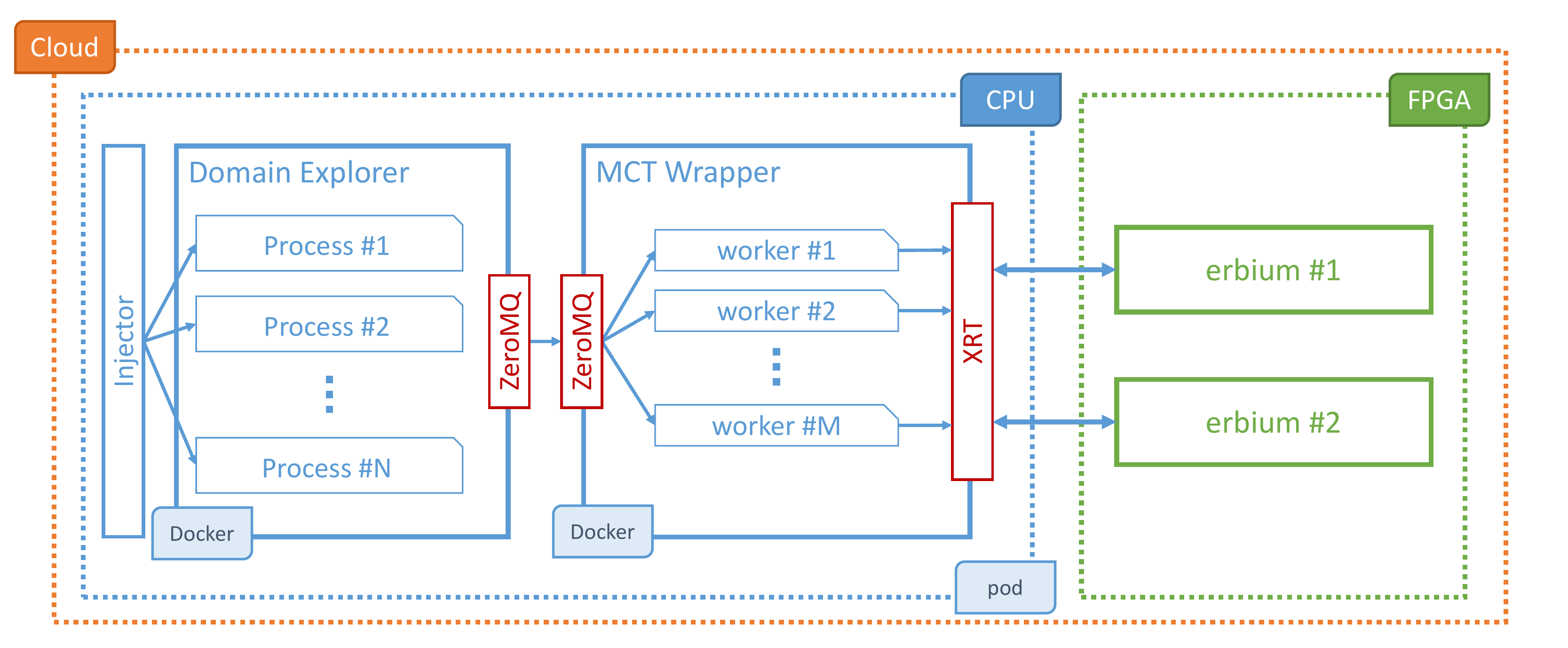}
  \caption{System Integration setup.}
  \label{fig:sysint_setup}
\end{figure}

We have ported the new version of \textsc{erbium} described in Section~\ref{sec:hardwaredesign} to a deployment of the Amadeus flight search engine on AWS F1 instances. The MCT queries used to generate load in the system are actual user queries captured from the production environment. The structure of the system is illustrated in Figure~\ref{fig:sysint_setup} and its main components are as follows:

\paragraph{Injector} The injector module generates the workload sent to the system by replaying the user query traces captured from the production environment. It distributes the workload to each one of the \emph{Domain Explorer} processes, saturating their processing capacity as much as possible.

\paragraph{Domain Explorer} A user query submitted to the search engine is executed by a single \emph{Domain Explorer} process, similarly to a pool of processes in a data management system; each process issues several \emph{Travel Solutions}, which then may trigger the evaluation of one to five MCT queries each. The number of processes that an instance of the \emph{Domain Explorer} deploys in one machine depends on the memory and number of cores available. In the current production environment, there are 48 \emph{Domain Explorer} processes in each machine.

\paragraph{ZeroMQ} The communication between the \emph{Domain Explorer} software with the new \emph{MCT Wrapper} uses the ZeroMQ framework. ZeroMQ is a light-weight networking library that handles concurrent communications using different patterns and protocols. It was already validated and deployed for similar purposes in other Amadeus applications, which facilitated its adoption in this case. We use a Request-Reply pattern combined with synchronous communications between the \emph{Domain Explorer} processes and the router, and asynchronous communications between the router and the multiple workers (\emph{MCT Wrapper} threads).

\paragraph{MCT Wrapper} The wrapper is a multi-threaded version of the \textsc{erbium} \emph{Host Executor} (Figure~\ref{fig:erbium}). It distributes the incoming MCT queries in a round-robin manner among the different workers using ZeroMQ \textit{dealers} to support asynchronous communication and, thus, making sure the wrapper is always ready to process new queries. The wrapper also plays an important role in hiding the details of the FPGA management form the rest of the system. 
The idea is to leverage a micro-service architecture to allow a fine-grained management of the deployment and the resources. Since the libraries used to communicate via PCIe to the FPGA are vendor-specific, using the wrapper avoids that the engine is tightly coupled to one particular vendor's set of tools or boards. A change from Xilinx to Intel FPGA boards would be encapsulated within the \emph{MCT Wrapper} (i.e., a new docker image), a transparent change vis-à-vis the \emph{Domain Explorer} software component. Following the same principle, this wrapper is also responsible for handling \textsc{erbium}-specific start up and update transactions, such as programming the FPGA with the generated bitstream and loading the NFA into the FPGA internal memory.


\paragraph{Encoder} The current \textsc{erbium} implementation uses dictionary encoding to reduce both the storage requirement and the online data movement. Therefore, the MCT query must be encoded before being sent to the accelerator. This process is carried out individually at the worker level in a pipeline manner, while the previous query batch is being processed by the \emph{erbium} FPGA kernel. This module is important in that it adapts data representations used in software to formats more amenable to FPGA processing. Without it, the design on the FPGA would not be competitive since it would be mired in the overhead of dealing with complex data types.

\paragraph{XRT} The communication between the different workers and the FPGA kernels is managed by the Xilinx Runtime Library (XRT). It schedules the data movement and the execution of the requests so that while the kernel is executing a batch of MCT queries, a different thread is being served by transferring its query data into the FPGA internal memory. We leverage this mechanism by having at least two threads per FPGA kernel.

\paragraph{FPGA} There are different possible combinations to fully utilise the FPGA resources. One can have, for instance, a single \textsc{erbium} kernel with four parallel engines; or two kernels with two parallel engines each. While the first setup is tuned to process a batch of queries as fast as possible (hence improving latency), the second prioritises the parallel execution of two batch of queries (thereby improving throughput).

\paragraph{Virtualisation} The individual instances of the \emph{Domain Explorer} and \emph{MCT Wrapper} run each on their own Docker container. All combined, they are packed into a Kubernetes pod, which is then associated to one FPGA board. Therefore, when a server provides multiple FPGA boards, there will be as many of these pods as there are boards. By deploying ZeroMQ clients and servers inside the same Kubernetes pod, we take advantage of Inter-process Communication (IPC) to minimise the communication overhead between containers. Depending on future needs, we could also deploy the Docker containers in different pods to better distribute the application across different machines. In that case, ZeroMQ would hide the change of communication protocol without impacting the rest of the system.

There is a 1-to-N relationship between the \emph{MCT Wrapper} and the FPGA board, so one board cannot be accessed by multiple \emph{MCT Wrappers}. It is both a technical limitation of the existing software stack provided by the FPGA vendors, where multiple containers (or pods) cannot be dynamically assigned to the same board, as well as a way to maximise the throughput by optimising the communication over PCIe, which is often the main bottleneck.

\subsection{System Overhead Characterisation}

\begin{figure*}
    \centering
    \begin{subfigure}[t]{0.50\textwidth}
        \centering
        \includegraphics[width=\linewidth]{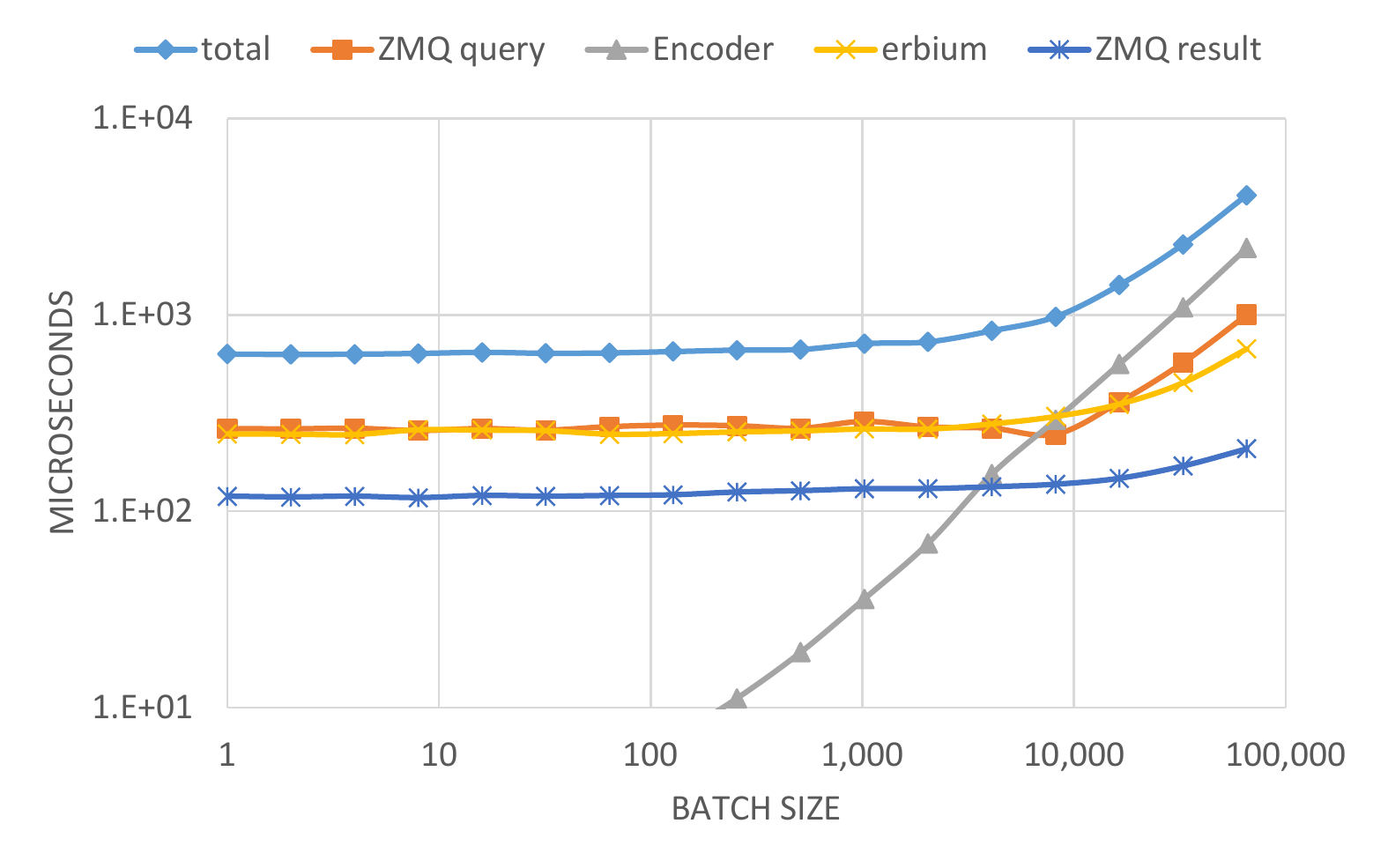}
    \end{subfigure}
    ~ 
    \begin{subfigure}[t]{0.50\textwidth}
        \centering
        \includegraphics[width=\linewidth]{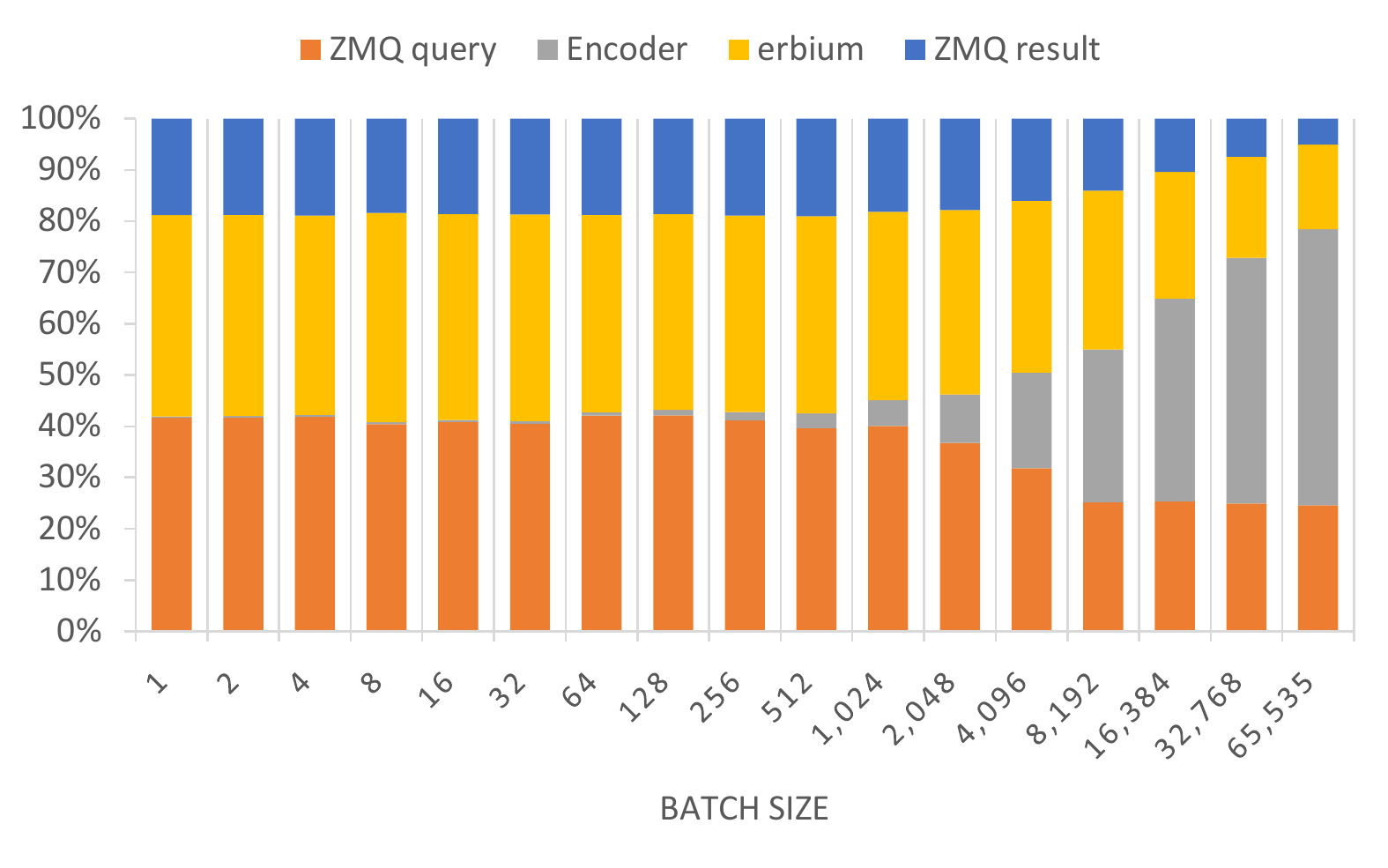}
    \end{subfigure}
    ~
    \caption{Execution Time of a MCT query decomposed into the different processing steps.}
    \label{fig:breakdown_responsetime}
\end{figure*}

The setup depicted in Figure~\ref{fig:sysint_setup} sheds light on the new execution elements required before and after the query evaluation by the FPGA kernel. Most of them can be parallised, such as the number of process within the \emph{Domain Explorer}, the number of workers within the \emph{MCT Wrapper}, the number of kernels within the FPGA, and the number of engines within the kernel. We first conduct a performance evaluation of the most basic scenario, where each one of those parallel elements is reduced to one: a single process generates the MCT queries, a single worker encodes the data and handles the communication with the FPGA, where a single \textsc{erbium} kernel evaluates them. With this configuration, we measure the intrinsic response time of each element processing different sizes of batches of MCT queries.

Figure~\ref{fig:breakdown_responsetime} shows the communication and synchronisation overheads of the different layers of the architecture as a function of the batch size. As a consequence of the PCIe bus constraints and FPGA shell protocols, the actual processing time for small batch size up to 4,096 MCT queries is dominated by data movement overheads. For larger batch sizes, the encoder imposes a linear and very high execution time, even bigger than the actual MCT query processing by the FPGA kernel. ZeroMQ communication overheads are also important, representing between 60\% to 30\% of the total response time for both query and response data movements. Given that the response time of the kernel execution on the FPGA is in the microsecond range, it is not a surprise that data encoding and data movements become equally or more expensive, a known problem in the context of data centre applications~\cite{BarrosoMPR17}. This is an important aspect to consider in view of the previous discussions about performance vs. easy of maintenance. In these ranges, and as the results show, the actual contribution to latency by the FPGA itself might not be the biggest issue in the overall design, removing some of the pressure to come up with the utmost optimised design. 


\subsection{Parallel Evaluation}

The system setup (Figure~\ref{fig:sysint_setup}) is very parallelisable. As a result, the deployment choices for each component are many, but may generate workload imbalances within the system. Globally, the correct choice of each parameter is the one that leads to a maximum overall throughput, while respecting the latency threshold and not under-utilising the system. We measure the global MCT throughput of the system as MCT queries per second, as well as the execution time of a user request (a batch of MCT queries) as seen by the \emph{Injector}. To measure the individual contribution of each parallel component, we conduct four series of experiments.

\begin{figure*}
    \centering
    \begin{subfigure}[t]{0.50\textwidth}
        \centering
        \includegraphics[width=\linewidth]{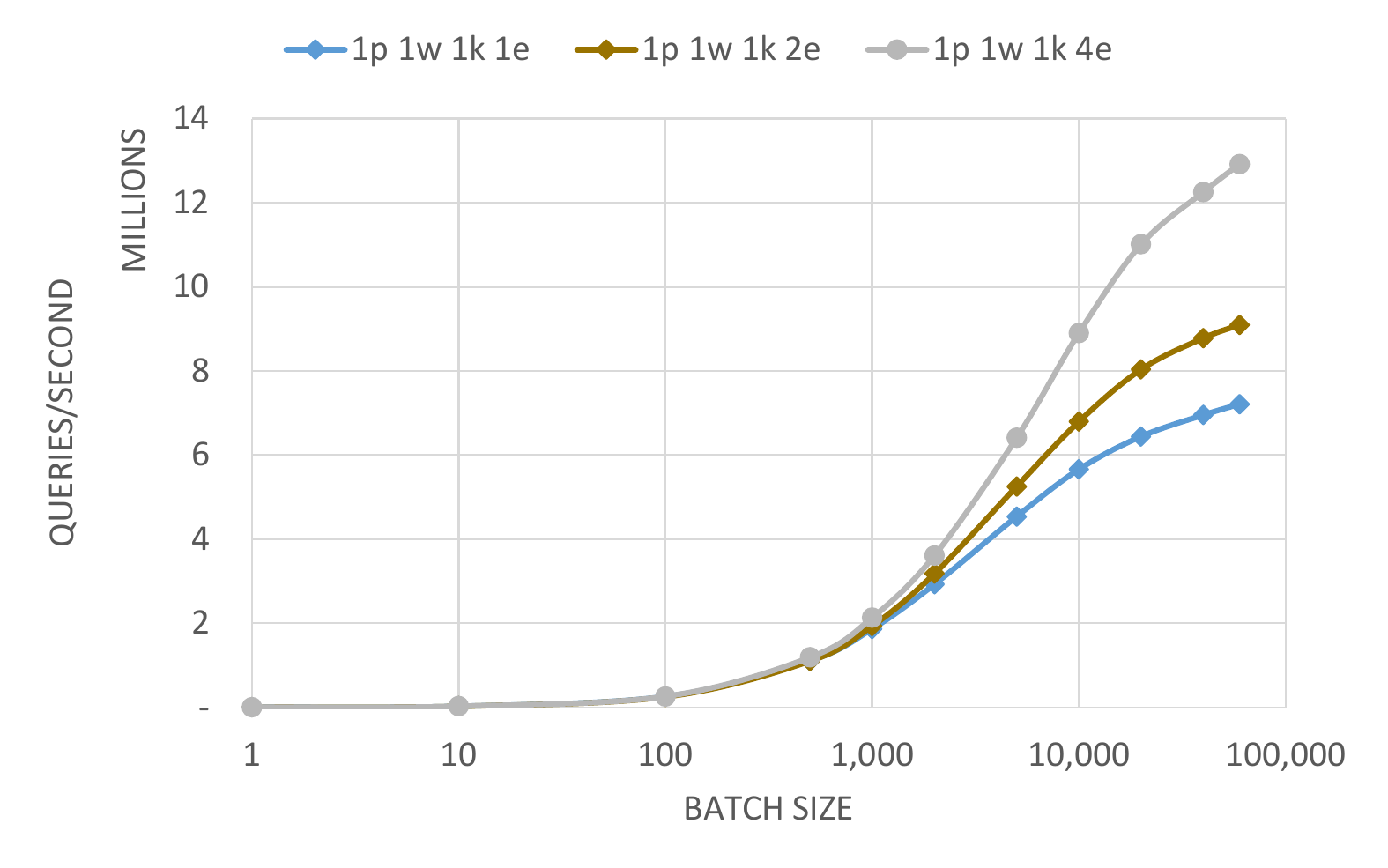}
        \caption{Global throughput in MCT queries per second}
        \label{fig:injection_engine_throughput}
    \end{subfigure}
    ~ 
    \begin{subfigure}[t]{0.50\textwidth}
        \centering
        \includegraphics[width=\linewidth]{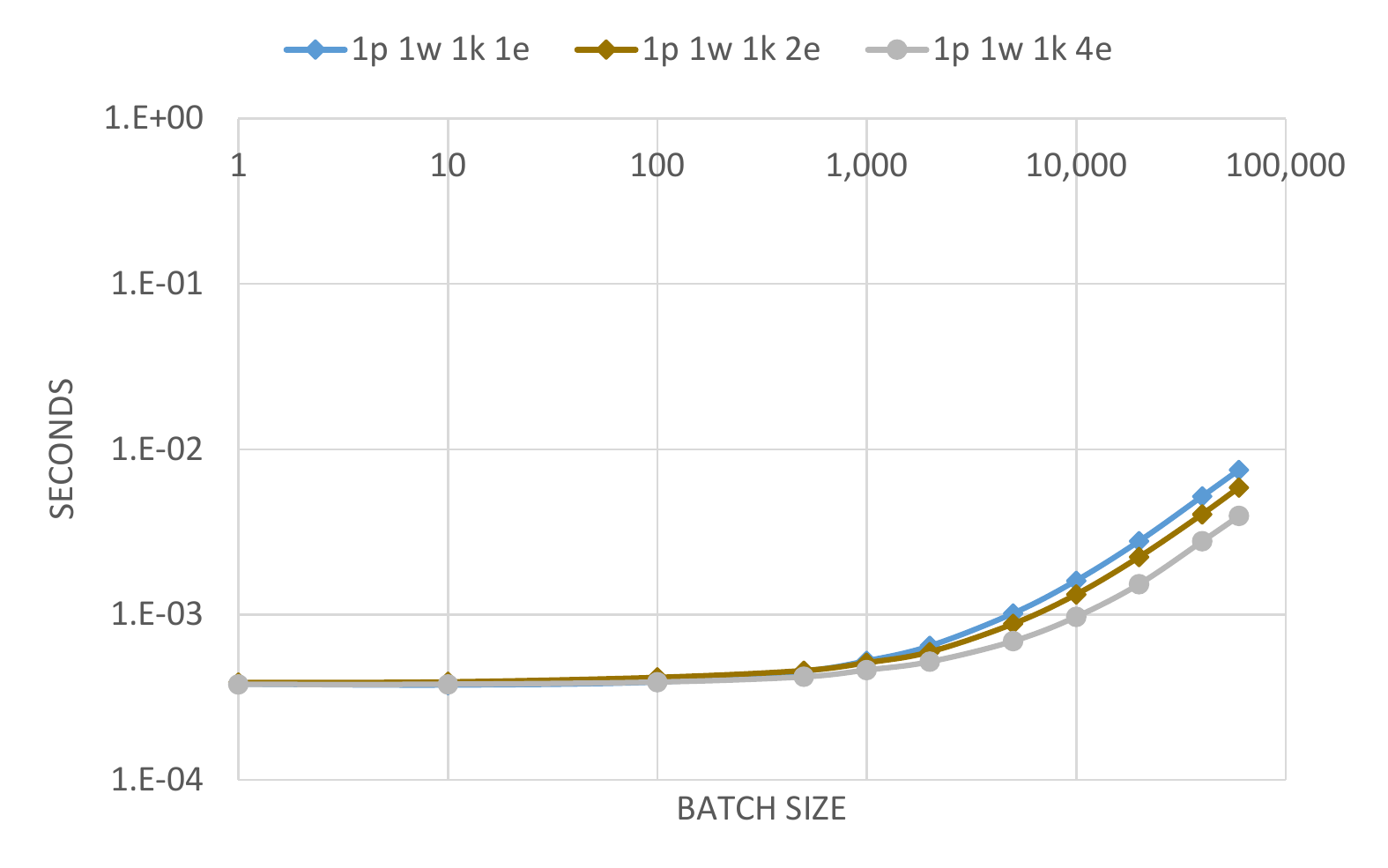}
        \caption{Execution time of a single MCT request}
        \label{fig:injection_engine_latency}
    \end{subfigure}
    ~
    \caption{Varying the number of engines per kernel. Series are labelled according to the number of processes (p), number of workers (w), number of kernels (k) and number of engines (e) per kernel.}
    \label{fig:injection_engine}
\end{figure*}

\begin{figure*}
    \centering
    \begin{subfigure}[t]{0.50\textwidth}
        \centering
        \includegraphics[width=\linewidth]{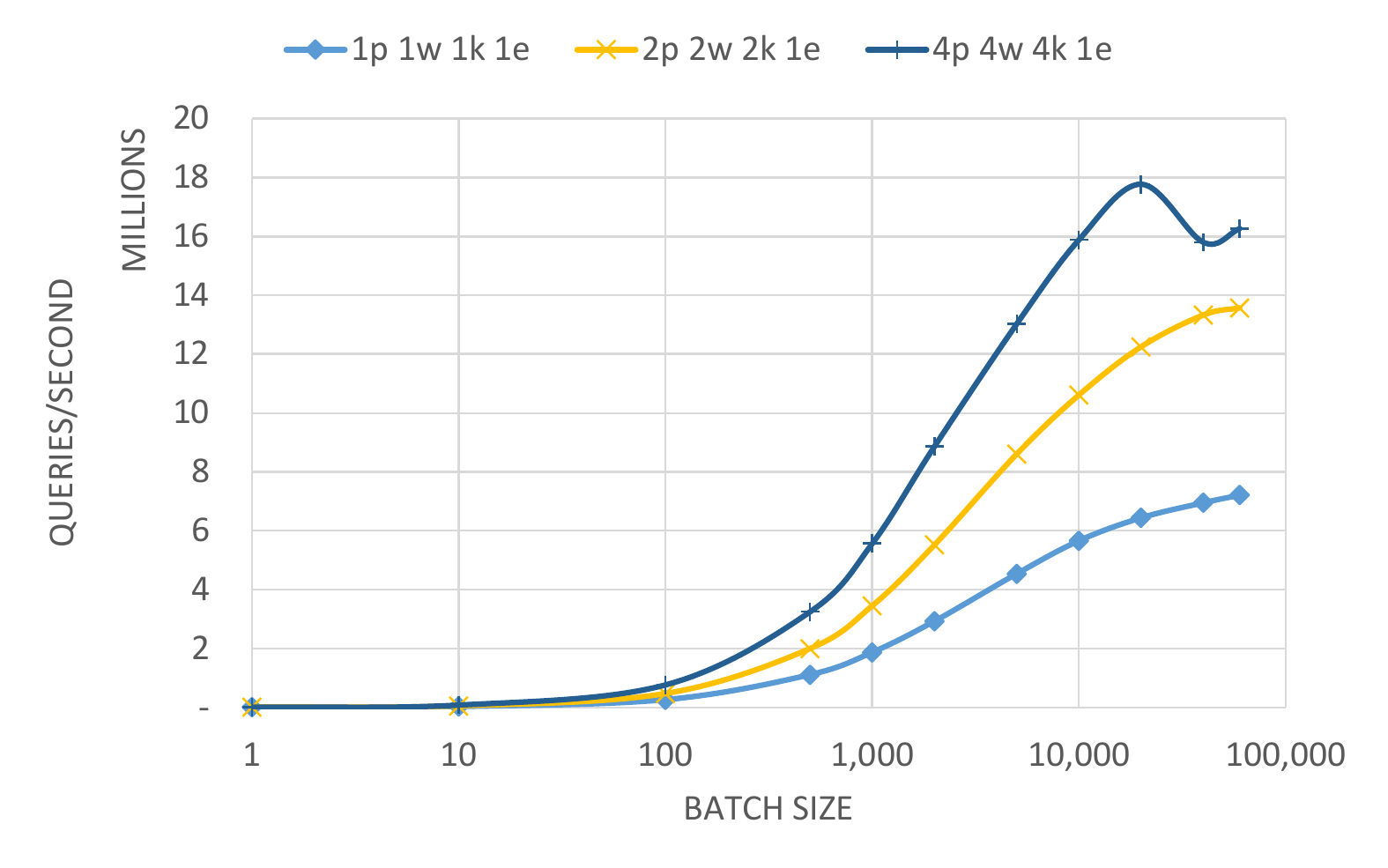}
        \caption{Global throughput in MCT queries per second}
    \end{subfigure}
    ~
    \begin{subfigure}[t]{0.50\textwidth}
        \centering
        \includegraphics[width=\linewidth]{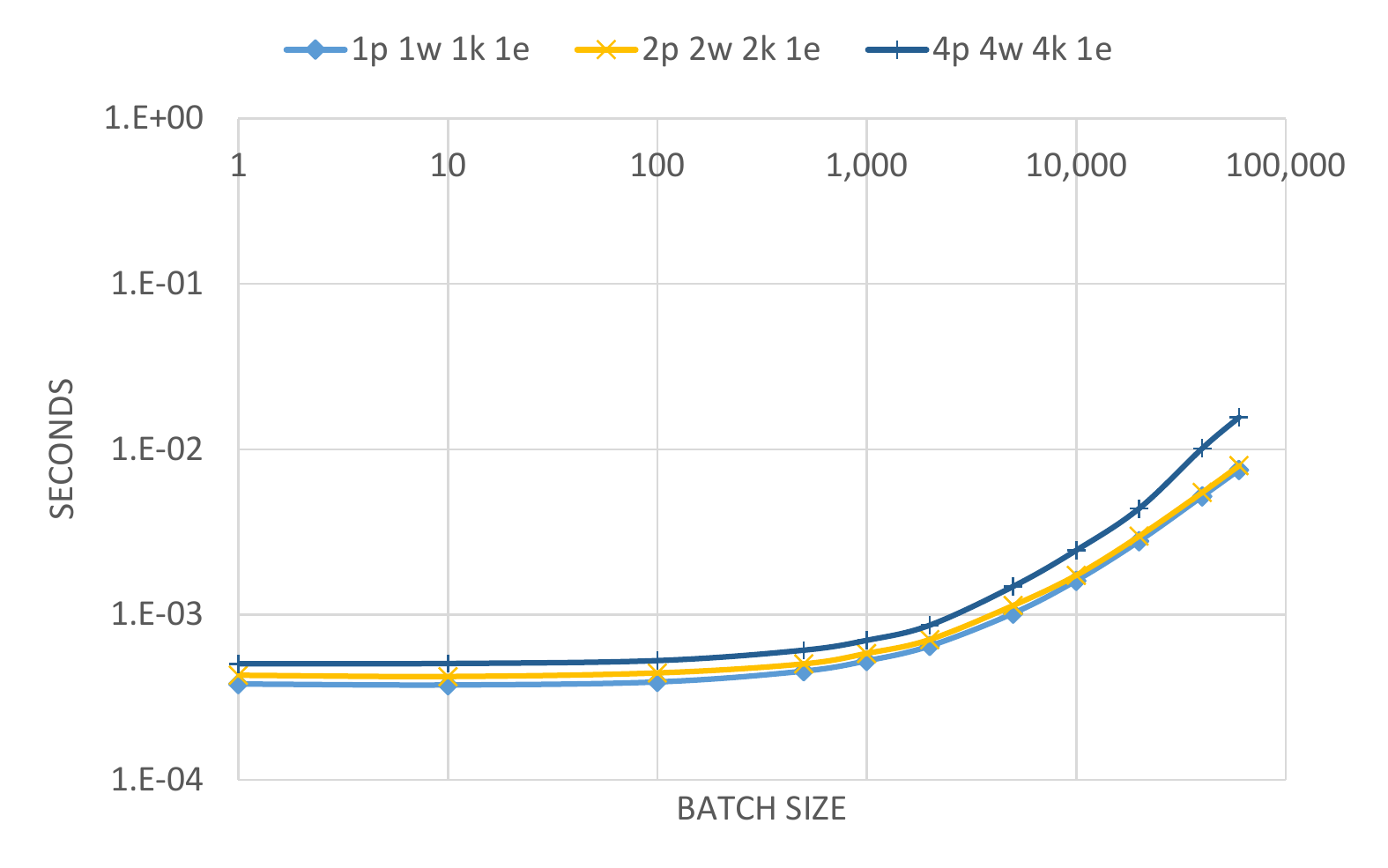}
        \caption{Execution time of a single MCT request}
    \end{subfigure}
    ~
    \caption{Varying the number of parallel components uniformly. Series are labelled according to the number of processes (p), number of workers (w), number of kernels (k) and number of engines (e) per kernel.}
    \label{fig:injection_linear}
\end{figure*}

\paragraph{Varying the number of engines per kernel} The available resources on a given FPGA board is fixed. The number of kernels that can be put into a single FPGA depends on the number of engines each one of them has. In the MCT v2 use case in the cloud, for example, the FPGA board is able to fit a total of 4 engines, so either one kernel of four engines, or two kernels of two engines, or four kernels of one engine each. We therefore measure the individual throughput gain for a single kernel varying the number of engines inside it for a fixed process and worker. In this setup, the execution of a single MCT request (a batch of MCT queries) is prioritised at the cost of a lower global throughput.

Figure~\ref{fig:injection_engine} presents the global throughput in MCT queries per second of kernels using one, two and four NFA Evaluation Engines. In this setup, a kernel distributes the workload of a single MCT request to all available engines. By increasing the number of engines, the request execution time is reduced, as shown in Figure~\ref{fig:injection_engine_latency}. In a single-process, single worker setup, this is also seen as a throughput increase. The circuit complexity increases with the number of engines, inducing a 30\% lower operating frequency. As a result, the performance numbers do not scale linearly to the number of engines.

\paragraph{Varying the number of parallel components uniformly} The second experiment is a complementary analysis of the previous compromise between number of engines per kernels and throughput vs. latency. We fix the number of engines per kernel, and measure the performance of adding new kernels. Each kernel has its own feeding flow (one process and one worker each). The system capacity to process parallel requests increases, as expected, as we add parallel processing elements (Figure~\ref{fig:injection_linear}). On the other hand, the latency of processing a single request increases, as the complexity of the FPGA circuit induces a slower operating frequency. In this setup, the global throughput is prioritised at the cost of a slower execution time per request (compare Figures \ref{fig:injection_engine} with lower throughput and lower latency and \ref{fig:injection_linear} with higher throughput but also higher latency).

\begin{figure*}
    \centering
    \begin{subfigure}[t]{0.50\textwidth}
        \centering
        \includegraphics[width=\linewidth]{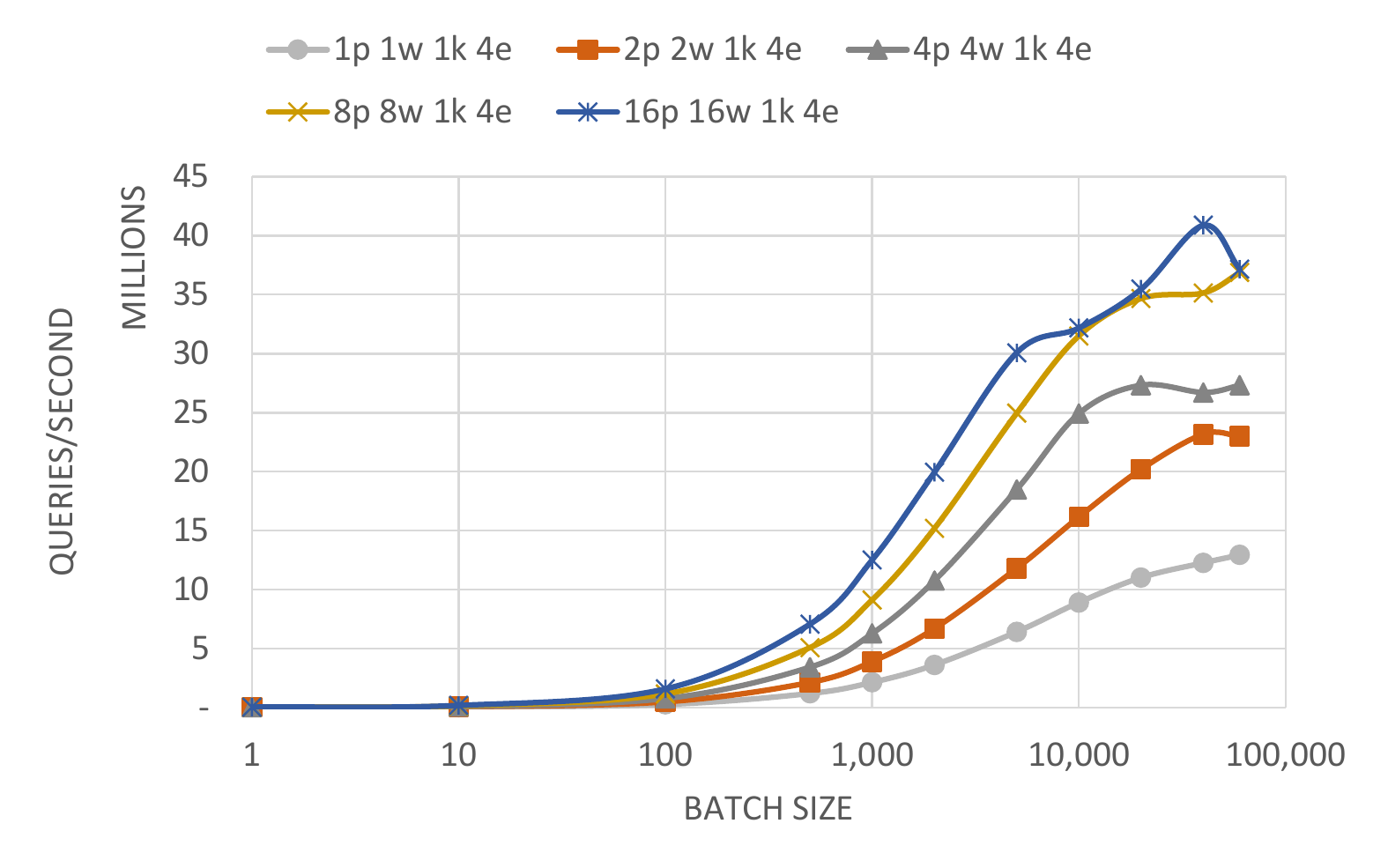}
        \caption{Global throughput in MCT queries per second}
    \end{subfigure}
    ~
    \begin{subfigure}[t]{0.50\textwidth}
        \centering
        \includegraphics[width=\linewidth]{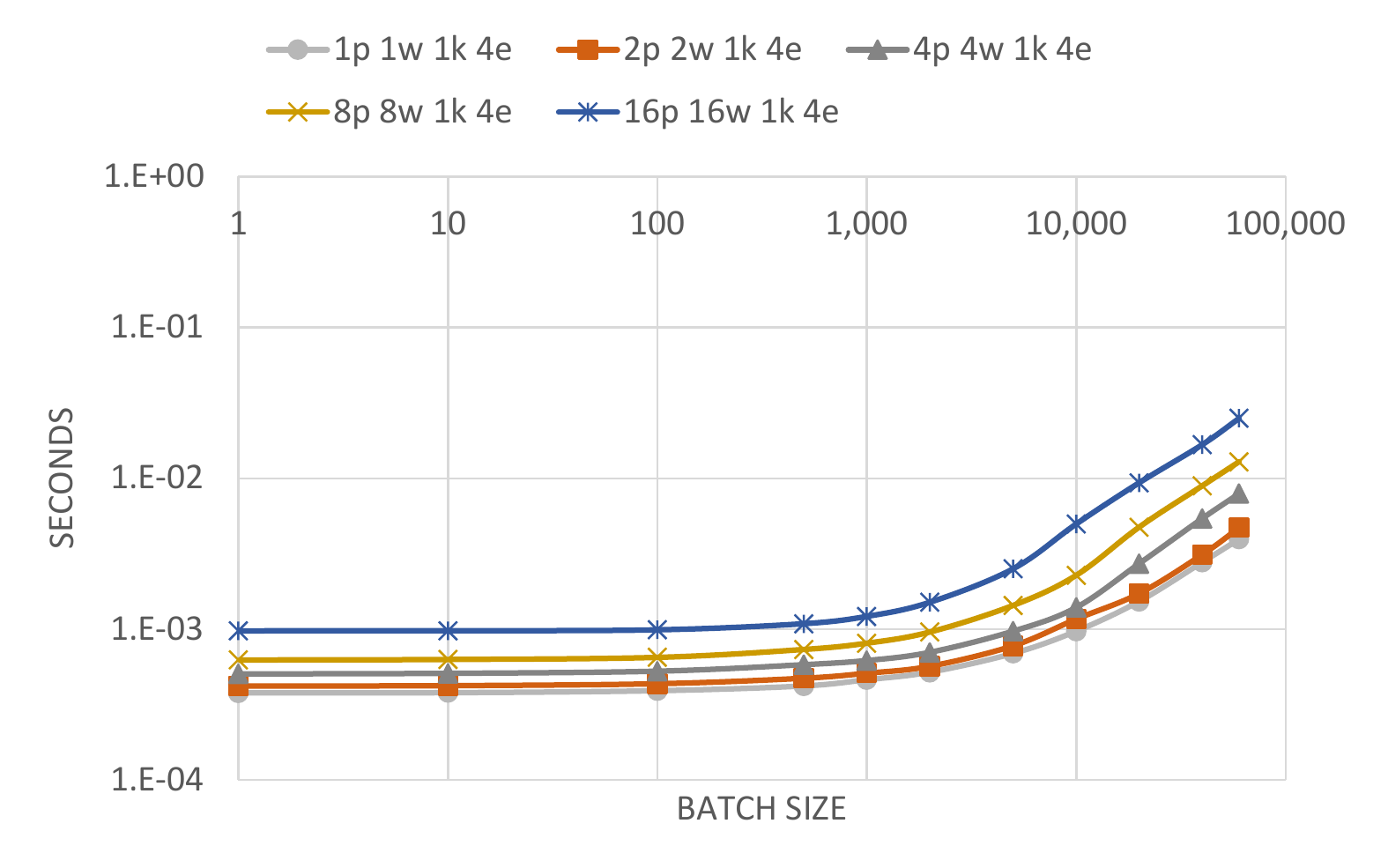}
        \caption{Execution time of a single MCT request}
    \end{subfigure}
    ~
    \caption{Multiple Process-Worker couple for a single kernel. Series are labelled according to the number of processes (p), number of workers (w), number of kernels (k) and number of engines (e) per kernel.}
    \label{fig:injection_couple}
\end{figure*}

\begin{figure*}
    \centering
    \begin{subfigure}[t]{0.50\textwidth}
        \centering
        \includegraphics[width=\linewidth]{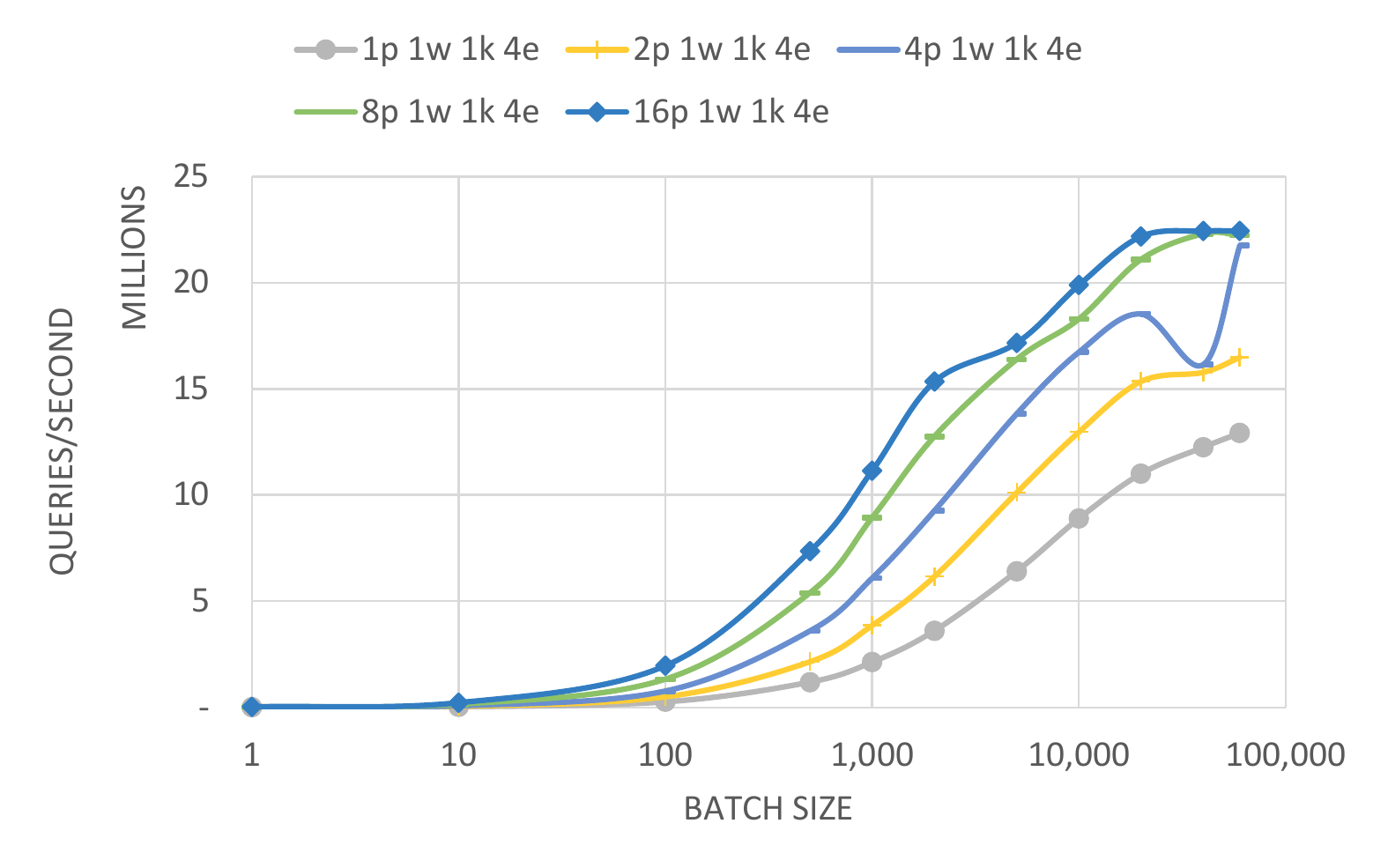}
        \caption{Global throughput in MCT queries per second}
    \end{subfigure}
    ~
    \begin{subfigure}[t]{0.50\textwidth}
        \centering
        \includegraphics[width=\linewidth]{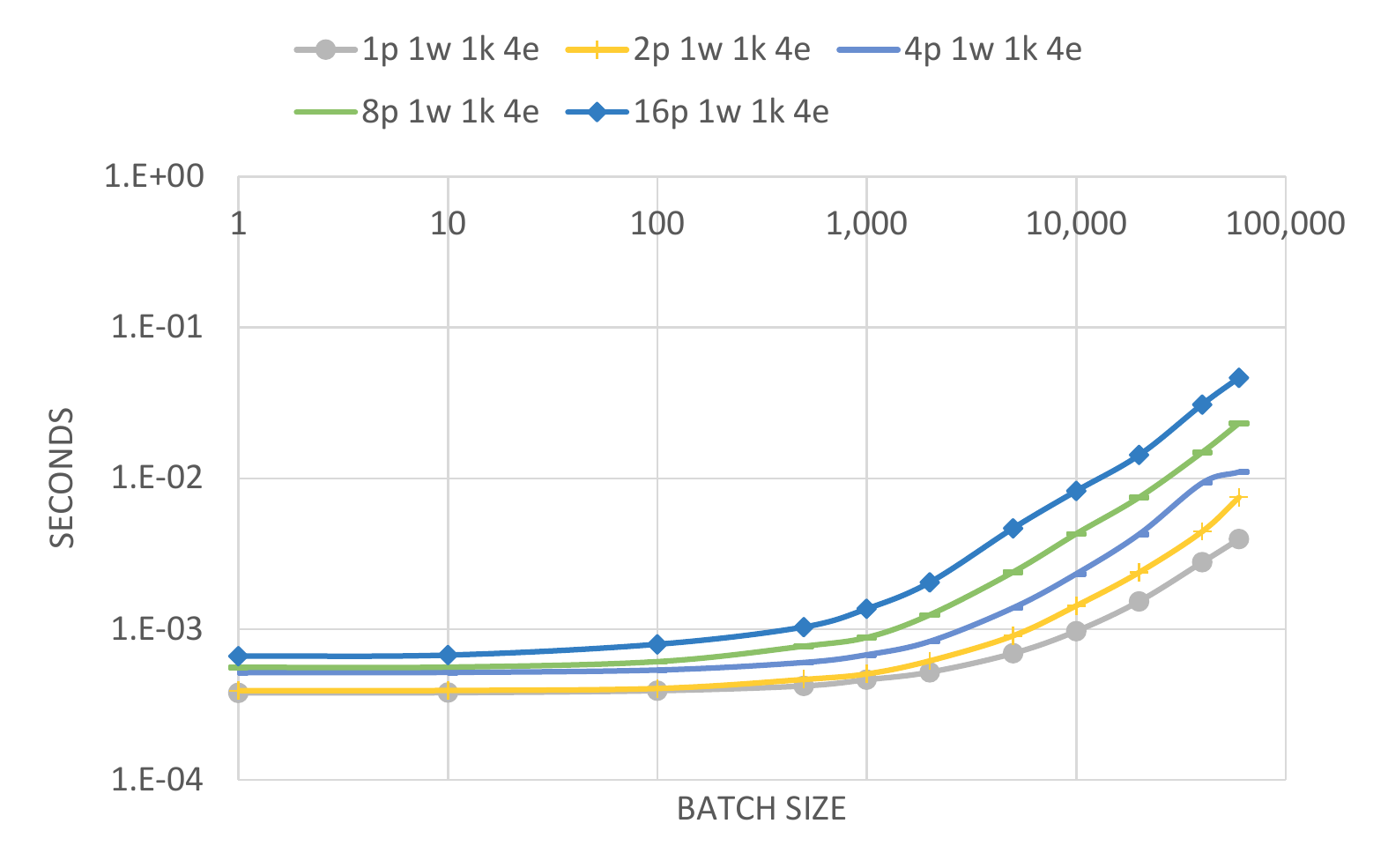}
        \caption{Execution time of a single MCT request}
        \label{fig:injection_process_latency}
    \end{subfigure}
    ~
    \caption{Multiple processes per worker. Series are labelled according to the number of processes (p), number of workers (w), number of kernels (k) and number of engines (e) per kernel.}
    \label{fig:injection_process}
\end{figure*}

\paragraph{Multiple Process-Worker couple for a single kernel} In this series of experiments, we measure how much workload a single kernel is able to handle as we increase the feeding flow. Given that an FPGA kernel stand-alone outperforms the global throughput (as shown in Figure~\ref{fig:mctv2benchmark}), the stress test of this experiment focuses on how much synchronisation overhead the \emph{XRT} driver imposes, since it schedules different MCT requests from different workers down to a single kernel. Figure~\ref{fig:injection_couple} shows that such a configuration maximises the global throughput, reaching up to 40 million MCT queries per second. On the other hand, the synchronisation overhead at the \emph{XRT} scheduler imposes a linear latency according to the number of feeding threads, but constant as regard to the batch size.

\paragraph{Multiple processes per worker} The final series of experiments focuses on stressing the workers at the \emph{MCT Wrapper}. We vary the number of processes a single worker is fed from. In this scenario, the worker is responsible for scheduling different MCT requests and batching them into a single \textsc{erbium} call. The worker then partitions the results and distributes them to the respective \emph{Domain Explorer} processes. We fix the kernel configuration as the best-case scenario from the first experiment (Figure~\ref{fig:injection_engine}), with four engines per kernel. Figure~\ref{fig:injection_process} shows that a single worker is not saturated by a single process, and therefore can deliver a bigger throughput when coupled to several processes, for example when connecting from 2 to 8 process per worker. The gain is reduced as we approach 16 process per worker, indicating a saturation at the worker level. In terms of execution time (Figure~\ref{fig:injection_process_latency}), the scheduling at the worker level imposes a similar latency than the one imposed at the \emph{XRT} level, but the former does depend on the batch size.

\subsection{Discussion}

\begin{figure}
  \centering
  \includegraphics[width=0.6\linewidth]{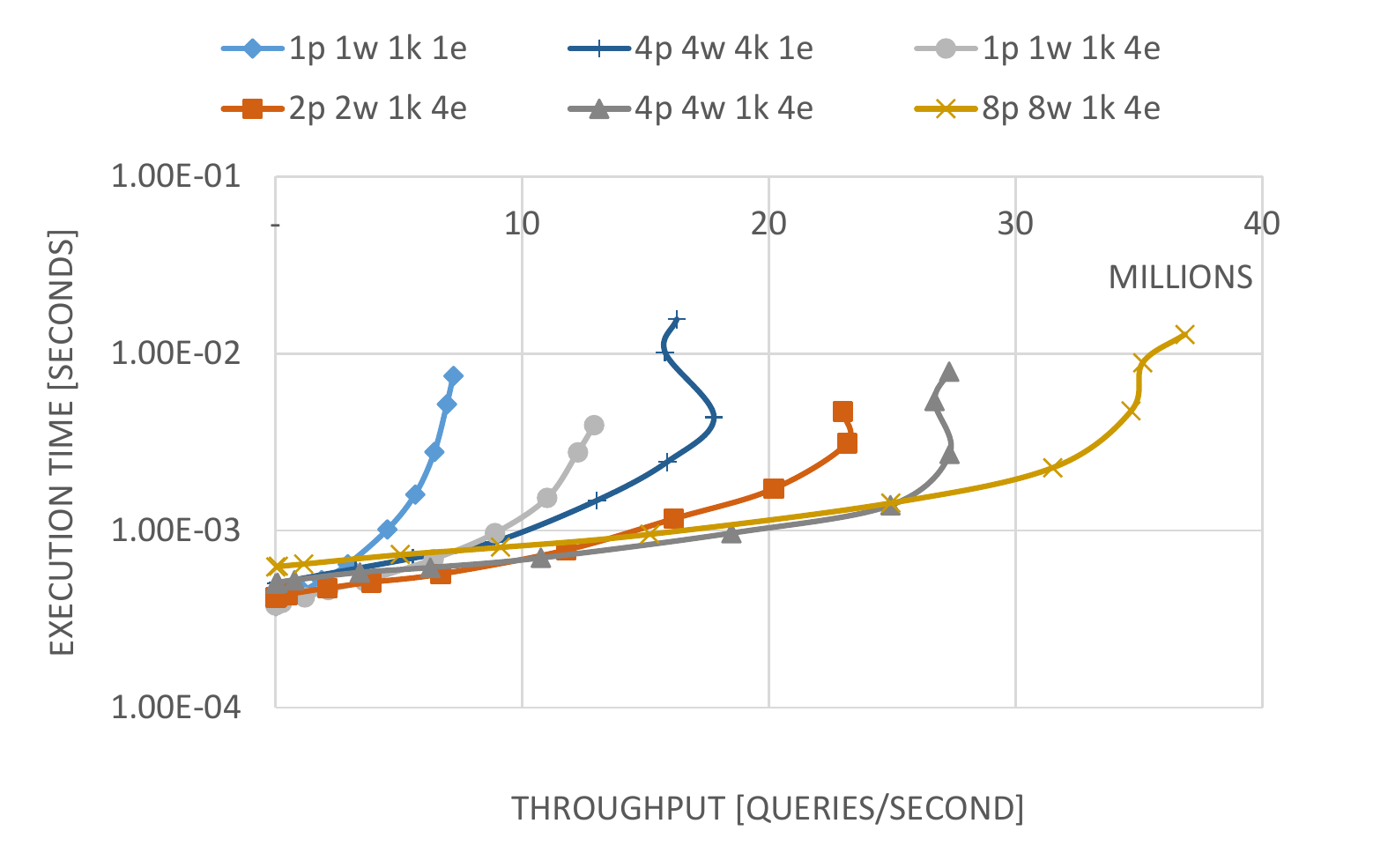}
  \caption{A pareto comparison of execution time as a function of the throughput for selected configurations. Series are labelled according to the number of processes (p), number of workers (w), number of kernels (k) and number of engines (e) per kernel.}
  \label{fig:injection_pareto}
\end{figure}

The choice of how many parallel elements are instantiated at each component of the Search Engine must take into consideration not only the direct gain in throughput that an additional parallel processing element introduces, but also how this element is coupled to the remaining ones~\cite{DBLP:journals/cea/Mazon-OlivoHMP18, chen2016spark}. The different configurations stress different components of the system, which present different overhead constraints. Global throughput can be maximised, at the cost of a higher execution time per user request. Contrary, fast response times per query lead to a lower global throughput. Figure~\ref{fig:injection_pareto} presents the compromise between these two dimensions and indicates which would be the best configuration according to the performance priority. For a given minimal throughput, e.g., 20 million queries per second, one can identify that a configuration with \texttt{4p 4w 1k 4e} would impose the lowest execution time; however if one fix the maximum execution time at, e.g., 500~$\mu\,s$, the configuration with \texttt{2p 2w 1k 4e} would be the one which yields the best throughput. Having such a flexibility in the configuration of the system and the degrees of parallelism is crucial in a search engine like the one considered here. Additional performance demands, whether higher throughput or lower latency, can be addressed by choosing the right combination of parameters. This flexibility also helps in avoiding over-provisioning by adjusting the size of the system to the target performance and in determining what element to scale out when needed.

From the research and practitioners perspective, an important lesson from these results is that the overall performance depends not only on the FPGA design, but also on the external elements of the system. Figure~\ref{fig:injection_pareto} shows a surprisingly large range of performance options both in latency and throughput, many of them determined by the system around the FPGA rather than by the design on the FPGA itself. Combined with the discussions above, these results confirm that absolute performance on the FPGA is less a factor than flexibility, easy of maintenance and evolution, and its integration within the rest of the system.

\section{Business Logic Analysis}
\label{sec:business_logic}
In this section we explore the impact that a higher level abstraction (the \emph{Travel Solution}) used in the search engine has on the MCT acceleration. Because the abstraction is currently tightly integrated in the current software implementation, eliminating it to accommodate the needs of the FPGA design would require a major redesign of the entire engine, something that would make the cost of adopting FPGAs prohibitive.

\subsection{From Travel Solutions to MCT queries}

The flight search engine manages requests in terms of \emph{Travel Solutions} (TS). A list of potential TS's is generated by the \emph{Connection builder} inside the \emph{Domain Explorer} just before the MCT call. This list is first sorted following an internal heuristic. Then, the \emph{Domain Explorer} sequentially reads the list, and calls the MCT module for the TS's that are not direct flights. Given that a TS would have in general no more than four stop-overs, the MCT request would be composed of at most four MCT queries to be checked. As soon as the \emph{Domain Explorer} has identified 1,500 valid TS's, it finishes the execution and returns the list of valid TS's to the next Search Engine component.

The current implementation of the \emph{Domain Explorer} is optimised for a CPU architecture, where the notion of batch processing is not required. In fact, the CPU has more flexibility and suffers less communication overhead than any PCIe-based device. On the FPGA, a similar approach of calling the MCT kernel with less than 10 MCT queries would not take advantage of the hardware acceleration (see below), as it can be seen in previous results (Figures~\ref{fig:breakdown_responsetime}-\ref{fig:injection_process}). A better approach is therefore to aggregate a number of MCT queries from different TS's coming from the same user query. However, the number of TS's to be evaluated is not known beforehand, and therefore there is another trade-off to resolve: minimise the number of TS's to be evaluated (since only the first 1,500 are used), but maximising the number of MCT queries packed into the same batch, in such a manner that only a single FPGA call is required. Recall that all experimental results shown above indicate larger batches are needed for the FPGA design to reach the necessary performance.

\subsection{A CPU perspective of the acceleration}

To measure how efficient the current CPU implementation is regarding the evaluation of independent TS's, we conduct an experiment using the architecture from Section~\ref{sec:systemintegration} deploying a single \emph{Domain Explorer} process, a single \emph{MCT Wrapper}, and a single \textsc{erbium} kernel with four NFA Evaluation Engines. In this experiment, instead of imposing the MCT batch size, we rely on the workload of individual TS's generated by the search engine.

We collected a snapshot of the workload of the search engine from production. This snapshot encompasses 6,301 real user queries that generate together 5.8 million potential TS's and 4.8 million MCT queries. About 17\% of the TS's of this benchmark correspond to direct flights, and therefore do not generate MCT calls. For the other ones, they spawn an average of 1.24 MCT queries each.

We proceed with a baseline comparison between a pure CPU execution using the current algorithms and the FPGA flow. The CPU baseline is a brand new, refactored and optimised version tailored for the MCT v2 use case. The new CPU version introduces some of the CPU optimisations previously reported~\cite{maschi2020}, as well as some cache mechanisms for selected airports.

To determine the batch size used for the FPGA call, we use the number of required qualified TS's provided by the user query. If the user query generates less potential TS's than the required qualified TS's number, all of the potential ones are batched together. In the other cases, we have multiple batches of the size of the required qualified TS's. This is not an optimal choice, since it does not minimise the number of FPGA calls whenever there are more potential TS's than the required number.

\begin{figure}
  \centering
  \includegraphics[width=0.7\linewidth]{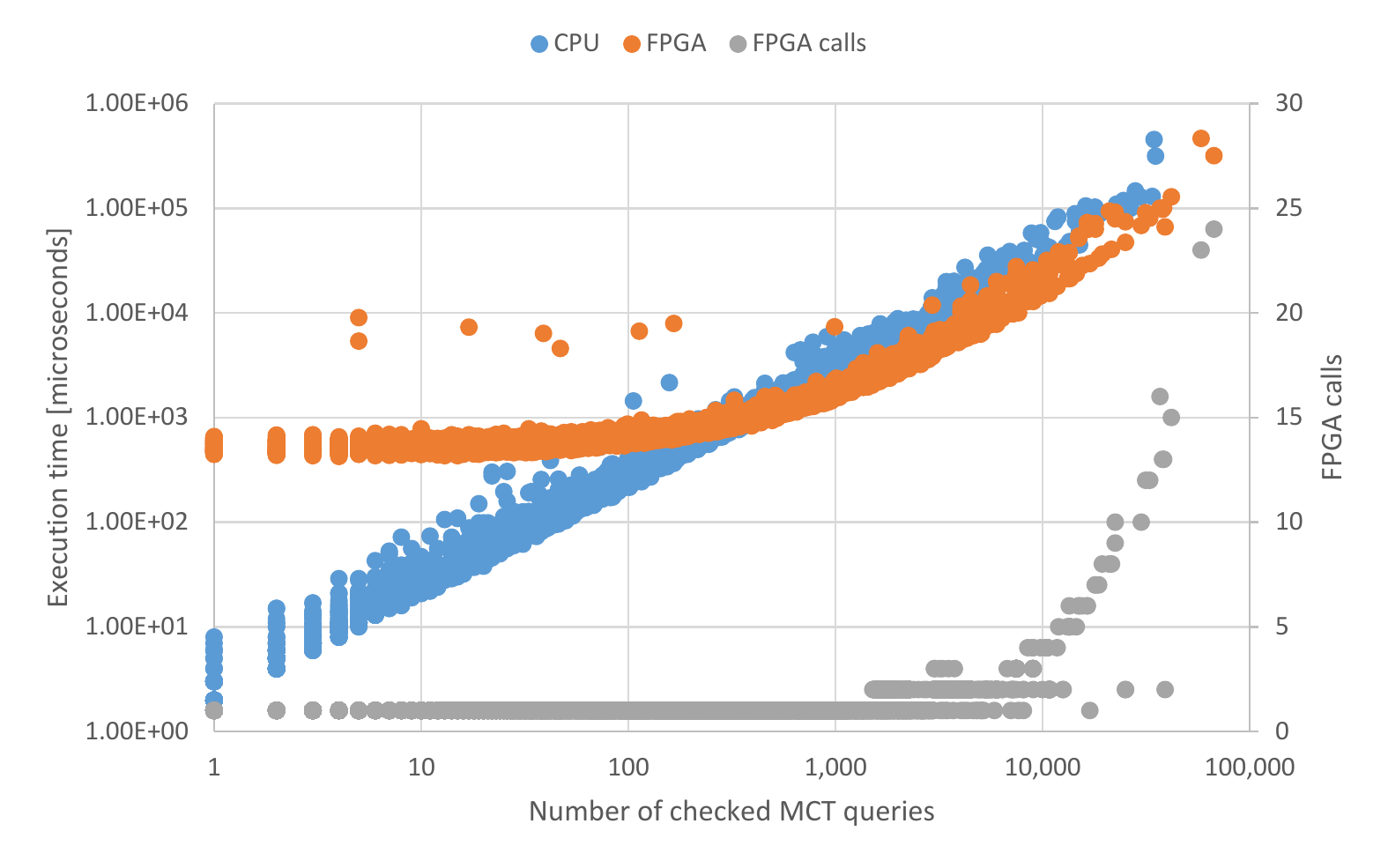}
  \caption{Execution time comparison of CPU and FPGA processing MCT queries. The number of FPGA calls to complete the request is also indicated.}
  \label{fig:sysint_perfcomp}
\end{figure}

Figure~\ref{fig:sysint_perfcomp} shows the execution time of individual user queries being processed by the CPU and FPGA MCT modules as a function of the number of checked MCT queries. We also plot the number of FPGA calls required to execute the user query. For small workloads of up to 400 MCT queries, the CPU implementations presents a smaller execution time, as the PCIe bus imposes an important communication overhead to the FPGA kernel. These small workloads could be packaged together, a solution that does not further penalise the execution time, but highly benefit the global throughput of the system, as shown in Figure~\ref{fig:injection_process}. For bigger workloads, the FPGA outperforms the CPU, even when called several times in the same request. 

\subsection{Discussion}

FPGAs are fundamentally different from CPUs and GPUs. While in all cases a certain amount of tuning to the underlying hardware makes sense, in the case of the FPGAs it is usually crucial to get the necessary performance. We have already seen this effect when considering the impact of the encoder in the overall overhead, which is necessary to adapt the incoming data types to formats the FPGA can process efficiently. In this section we have seen another aspect that plays a crucial role. All experimental results indicate that the batch size has to be large enough for the FPGA to be competitive. However, the batch size is dictated by the search engine and how it works. We have been able to find a compromise solution, but the overall result is that the performance on the FPGA suffers. At small batch sizes, there is not enough load for the FPGA to be fully exploited and the overhead of sending the requests to the FPGA dominates. The size of the batch that can be processed is determined by how the flight search engine deals with \emph{Travel Solutions}. We can delay submitting queries to batch several requests, but that has an impact on latency and requires additional logic before the FPGA. Moreover, not all user queries result in requests or enough requests to the MCT, naturally leading to small batch sizes and infra-utilisation of the FPGA. One could argue that the flight search engine could be modified to accommodate the FPGA. This is not a realistic option in the short term as it would be very costly and still does not address the fact that the load is not uniform: often enough there are not enough requests for the MCT module. These effects play no role in the CPU implementation, but turn the FPGA deployment into a rather complex compromise in terms of design and performance with the added factor that the FPGA cannot always be fully exploited.



\section{System Deployment on Current Cloud Platforms}
\label{sec:systemdeployment}

In this section we present a brief analysis of the cost of deploying the FPGA-based solution using \textsc{erbium} for both on-premises and in the cloud. These numbers are approximate, but they give an accurate picture of the issues involved. 

\subsection{\emph{Domain Explorer} and \textsc{erbium}}

\begin{figure}
  \centering
  \includegraphics[width=0.7\linewidth]{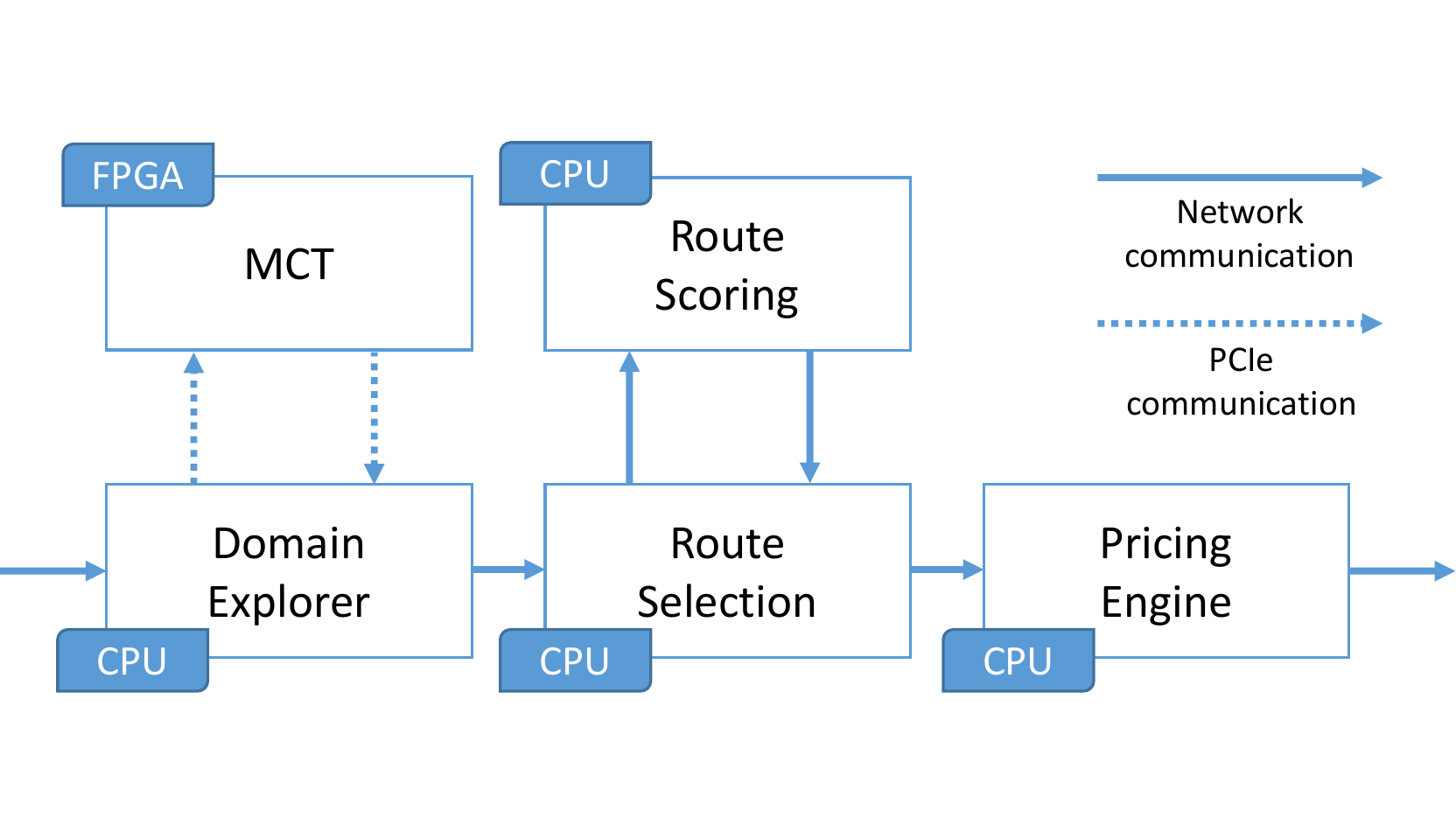}
  \caption{Moving the MCT module~\cite{maschi2020} to an FPGA PCIe-attached to the Domain Explorer.}
  \label{fig:pfx-erbium}
\end{figure}

The more direct implementation of the ideas proposed places the hardware accelerator as a co-processor attached to the CPU through PCIe (Figure~\ref{fig:pfx-erbium}). This setup is the only one available in both AWS F1 and Azure NP-series, as of February 2021. By running MCT on an FPGA, we could free up to 40\% of CPU time in the \emph{Domain Explorer}, which could then, theoretically, accept 40\% more requests. To avoid under-use of the computing resources, an alternative is to reduce the number of CPUs by 40\%, since that will still leave enough capacity with the current loads. 

The strategies to follow to take advantage of the new design are different whether the targeted deployment is on-premises or in the cloud. For on-premises, the search engine already exists, therefore the computing resources is already accounted for. An eventual re-allocation of resources within the company should also be considered, for instance, allocating the surplus CPUs to a different service or product. Moreover, for on-premises deployments, the choice of computing resources (e.g., FPGA board and CPU configuration) is driven by the cost and the technical maintenance overhead within the company. For deployments in the cloud, it is important to adjust the resources to the demand so under-utilisation should be kept to a minimum. The drawback of deployments in the cloud arises from the currently available configurations of CPU and FPGA models, FPGA shells (and hence, communication interfaces) and how they are attached together. As of February 2021, both Amazon F1 and Azure NP-series instances provide a single option: a big FPGA (UltraScale+ VCU9P and Alveo U250, respectively), PCIe-attached to a small CPU (8 and 10 vCPUs, respectively). This offer is a major handicap in a co-located service architecture, where the CPU has to run the \emph{Domain Explorer}. To have the same CPU capacity as one CPU-only server used on-premises, we would need about 6 AWS F1 instances, for example. Additionally, there is currently no possibility of multiples CPU instances sharing the same FPGA, or FPGA-to-FPGA communication. The result is that while the FPGA can process a large amount of MCT requests, the CPU side is simply not powerful enough to generate enough load.

\begin{table}
  \centering
  \caption{Rough cost estimate for different deployments on-premises and in the cloud for the \emph{Domain Explorer} and MCT modules configured as shown in Figure~\ref{fig:pfx-erbium}. Hourly prices for AWS instances in Europe (Ireland) region, and for Azure instances in United States (Washington), both for a savings plan of one year.}
  ~
  \begin{tabular}{@{}lllrrr|r@{}} \toprule
  & & Element & vCPUs & Units & Unit Cost & Total \\
  & & & & & (USD) & (USD) \\ \midrule
  \multicolumn{3}{l}{On-Premises} \\
  & Original \emph{Domain Explorer} & CPU & 48 & 400 & 10 k & 4 M \\
  & \emph{Domain Explorer} + \textsc{erbium} & CPU + Alveo U200 & 48 & 244 & 20 k & 4.88 M \\ 
  & \emph{Domain Explorer} + \textsc{erbium} & CPU + Alveo U50 & 48 & 244 & 13 k & 3.17 M \\ \midrule
  \multicolumn{3}{l}{AWS} \\
  & Original \emph{Domain Explorer} & c5.12xlarge & 48 & 400 & 1.452/h & 5.0 M/year \\
  & \emph{Domain Explorer} + \textsc{erbium} & f1.2xlarge & 8 & 1,464 & 1.2266/h & 15.7 M/year \\ \midrule
  \multicolumn{3}{l}{Azure} \\
  & Original \emph{Domain Explorer} & F48s v2 & 48 & 400 & 1.2084/h & 4.2 M/year \\
  & \emph{Domain Explorer} + \textsc{erbium} & NP10s & 10 & 1,171 & 1.0411/h & 10.6 M/year \\ 
  \bottomrule
  \end{tabular}
  \label{tab:financial_erbium}
\end{table}

Table~\ref{tab:financial_erbium} summarises a simplified calculation of the costs associated to different deployments based on the current load (for which 400 servers are needed) and the proportion of the load that the FPGAs could take over in the best case. We use 400 large multi-core servers at a purchase price of 10K each as a baseline. The equivalent in the cloud are \textit{c5.12xlarge} and \textit{F48s v2} instances, endowed with 48 vCPUs and 96~GiB of memory each. Assuming the FPGA takes over 40\% of the load, we only need then 244 large servers with an FPGA attached. In the cloud, because the CPU side of the available instances is so small, we would need 1,464 \textit{f1.2xlarge} or 1,171 \textit{NP10s} instances to be able to cope with the current load on the rest of the system.

This cost calculation is not intended to provide a full comparison between on-premises vs. cloud deployments, since it does not take into account maintenance, facility, energy and operational costs on-premises. This being said, it does provide a good base of comparison for the different system layouts among each category. Notably, these numbers demonstrate that, on-premises, the new design is only cost-effective when using a smaller FPGA, as the larger ones are too expensive. Especially considering that we cannot fully utilise the FPGA. However, the MCT \textsc{erbium} design would fit on an Alveo U50, which has a list price about a fourth of that of the larger boards. With such a board, the price of the system becomes more competitive. In the cloud, the current configurations are simply inadequate for the proposed design. The increase in cost~---~3x for AWS, and 2.5x for Azure~---~over a CPU-only design is prohibitive.

\subsection{\emph{Domain Explorer} and \textsc{erbium} and \emph{Route Scoring}}
The calculation above only considers the MCT module. As we have seen, several effects make it difficult to keep the FPGA at capacity. An alternative design is to add other modules of the search engine to the FPGA, so that it has enough load in spite of the caveats discussed so far. This could be done by combining the MCT module with the \emph{Route Scoring} module, a different part of the flight search engine successfully accelerated using FPGAs~\cite{Mohsen20}. 

\begin{figure}[t]
  \centering
  \includegraphics[width=0.7\linewidth]{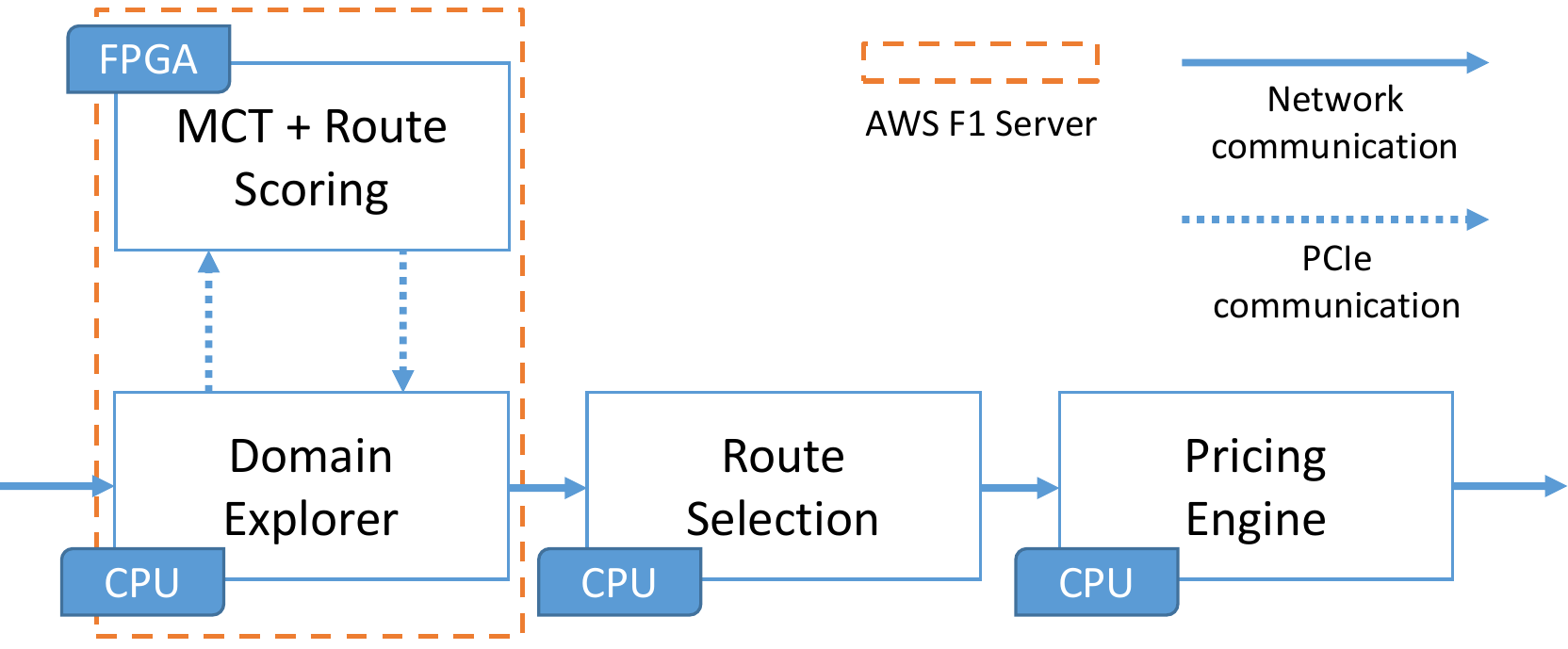}
  \caption{Moving \emph{Route Scoring}~\cite{Mohsen20} and MCT~\cite{maschi2020} to an FPGA running on the same machine as the Domain Explorer.}
  \label{fig:pfx-future}
\end{figure}

The \emph{Route Scoring} would move earlier in the flow, directly inside the \emph{Domain Explorer}, to score the routes during the flight domain exploration (Figure~\ref{fig:pfx-future}). By doing so, the \emph{Route Scoring} would be able to process several tens of thousands of routes in the \emph{Domain Explorer}, instead of only few hundreds inside the \emph{Route Selection}, while respecting the same response time constraint. Both MCT and \emph{Route Scoring} would then be on the same FPGA, and pipelined together to minimise the back and forth with the CPU, notably the PCIe overhead. Such pipeline would require an additional refactoring of the \emph{Domain Explorer}, since the notions of route for MCT and \emph{Route Scoring} are not exactly the same, but such deployment would optimise the FPGA occupancy and its usage, while combining the potential business benefits of both modules.

\begin{table}[b]
  \centering
  \caption{Rough cost estimate for different deployments on premise and on AWS instances for the \emph{Domain Explorer}, MCT and \emph{Route Scoring} modules configured as shown in Figure~\ref{fig:pfx-future}. Hourly prices for AWS instances in Europe (Ireland) region, and for Azure instances in United States (Washington), both for a savings plan of one year.}
  ~
  \begin{tabular}{@{}lllrr|r@{}} \toprule
  & & Element & Units & Unit Cost & Total \\
  & & & & (USD) & (USD) \\ \midrule
  \multicolumn{3}{l}{On-Premises} \\
  & Original \emph{Domain Explorer} + \emph{Route Scoring} & CPU & 480 & 10 k & 4.8 M \\
  & \emph{Domain Explorer} + \textsc{erbium} + \emph{Route Scoring} & CPU + Alveo U200 & 244 & 20 k & 4.88 M \\ 
  & \emph{Domain Explorer} + \textsc{erbium} + \emph{Route Scoring} & CPU + Alveo U50 & 244 & 13 k & 3.17 M \\ \midrule
  \multicolumn{3}{l}{AWS Instance} \\
  & Original \emph{Domain Explorer} + \emph{Route Scoring} & c5.12xlarge & 480 & 1.452/h & 6.1 M/year \\
  & \emph{Domain Explorer} + \textsc{erbium} + \emph{Route Scoring} & f1.2xlarge & 1,464 & 1.2266/h & 15.7 M/year \\ \midrule
  \multicolumn{3}{l}{Azure} \\
  & Original \emph{Domain Explorer} + \emph{Route Scoring} & F48s v2 & 480 & 1.2084/h & 5.0 M/year \\
  & \emph{Domain Explorer} + \textsc{erbium} + \emph{Route Scoring} & NP10s & 1,171 & 1.0411/h & 10.6 M/year \\ 
  \bottomrule
  \end{tabular}
  \label{tab:financial_future}
\end{table}

Table \ref{tab:financial_future} contains a similar cost estimate to the previous one, but this time including the \emph{Route Scoring} module in the equation. To the 400 CPU-only servers needed by the \emph{Domain Explorer}, we now add 80 servers to run the \emph{Route Scoring} module. That pushes the on-premises cost of the CPU-only solution to 4.8 millions. Because now the FPGA is used to replace 40\% of the \emph{Domain Explorer} servers plus all of the \emph{Route Scoring} severs, the cost-effectiveness of an FPGA-based solution improves over the previous scenario. It is still the case that we would be able to run the two FPGA designs, MCT and \emph{Route Scoring}, on the Alveo U50; so we include the costs for that case as well (although we have not performed extensive experiments to find out the best possible configurations and how using both designs at the same time affects the PCIe bottleneck). In the cloud, the cost associated to the additional servers for \emph{Route Scoring} raise the total cost of the flight search engine, but not enough to bring it anywhere near what the cost would be for the FPGA-based solution given the current prices and configurations. One could argue that, both in this as well as in the previous scenario, there could be ways to use less F1 instances. For instance, in the first scenario, one could direct user queries for direct flights to instances without an FPGA and request requiring the MCT to F1 instances. In practice, this implies major changes to the engine, which take time and also incur a considerable cost. At the end, the difference in price is so large that it is unlikely any straightforward solution will make the FPGA-based deployment cost-effective in the cloud under the current computing resource offer layouts. Only a different configuration with a much more powerful CPU that would significantly lower the number of servers needed would solve the problem, assuming that the cost of the new instance is not much higher than the current ones.

\subsection{Discussion}

Although these results are undoubtedly rough cost calculations and some of the prices of servers, boards, and cloud instances can vary in the near future, they do provide a sobering perspective on large scale FPGA deployments using today's systems. The main issue that these results bring to the fore is the large imbalance between the CPU and the FPGA in the cloud instances. We have shown throughout the paper that turning the initial prototype into a real system results in a performance loss over the ideal case, performance that is further reduced by impedance mismatches between the current system and what the FPGA needs to provide maximise performance. Add to that that the instances in the cloud do not provide enough computing capacity for the software part of the engine to put enough load on the FPGA, and the result is a prohibitive cost for FPGA deployments. The obvious solution to this is to have access to instances with a different configuration. On the research side, though, the results point out at the importance of considering how to put enough load on the FPGA and the overhead of getting results back. PCIe-attached boards hit the bandwidth bottleneck very quickly, especially if the shells do not support streaming. In previous work for the \emph{Route Scoring} module~\cite{Mohsen20} we considered such an overhead and found out that we lost a significant part of the advantage of using an FPGA, even if, in total, the FPGA-based solution was still ahead. 

A direct consequence of these ideas is that it might turn out that the better way to use FPGAs as accelerators is not as co-processors attached to the CPU, but as first class citizens that are accessible directly from the network. That could look like the configuration used in Microsoft Azure~\cite{Firestone2018}, where the FPGA is available in the network. Alternatively, one could think of having stand-alone FPGAs or FPGA clusters also directly connected to the network, as IBM has proposed on their CloudFPGA~\cite{CloudFPGA16,CloudFPGA19}. In the case of the \emph{Route Scoring} module, this is trivial to do, as the module is already a separate component running on a separate server. For the MCT module, things are more complex, as what is currently done with a local call becomes a network invocation, which will add to the latency. Nevertheless, one could think of placing all FPGA accelerated modules on the FPGAs as discussed above and change the search engine to use the module remotely. This is an idea that we intend to explore in the future using both TCP/IP communication~\cite{Limago}, as well as RDMA~\cite{strom}.




\section{Conclusion}

Data centres make computing heterogeneity affordable. While academic research has been showing great potential for the use of FPGA to accelerate applications, integrating them into complex and legacy systems is yet a challenge. In this paper we provide an extensive study on the steps necessary to move from a research prototype of an FPGA-based accelerator to a Proof-of-Concept integrated within a realistic system under realistic constraints. The first lesson is that the same trade-offs between flexibility, optimisation and maintainability met in the context of High Performance Computing software is also valid for FPGA-based accelerators. Although there has been a lot of research effort to maximise performance of FPGA designs at all costs, long-term deployment in systems that evolve over time may make such highly optimised designs too expensive to maintain. Moreover, we show that orders of magnitude better performance \emph{vis-à-vis} an independent function of the system may be significantly less as a result of mismatches between the software and hardware. In the context of legacy systems, such a co-design effort imposes an important development cost. Finally, the FPGA offerings currently available in the cloud are still too limited, preventing a full utilisation of the FPGA computing power by lack of CPU power. For on-premises deployment, however, the advantage of more freedom of choice on the hardware architecture makes FPGA designs more competitive. We hope that these insights will help and promote more research considering the end-to-end aspects of FPGA deployments and also inform existing deployments to expand the range of options available.

\begin{acks}
This paper extends and builds upon a previous publication by the authors that appeared in ACM SIGMOD 2020~\cite{maschi2020}. We would like to thank Xilinx for the generous donations of the Alveo boards and the support in the development of the Proof-of-Concept prototype. Part of the work of Fabio Maschi was funded by a grant from Amadeus. 
\end{acks}


\bibliographystyle{ACM-Reference-Format}
\bibliography{journal-trets21}

\end{document}